\documentclass[aps,prd,twocolumn,groupedaddress,nofootinbib]{revtex4-2}

\usepackage{graphics,graphicx}
\usepackage{amsmath, amssymb}
\usepackage{natbib,hyperref}
\usepackage{mathtools}
\usepackage{hyperref}

\DeclareMathOperator*{\sumint}{%
\mathchoice%
  {\ooalign{$\displaystyle\sum$\cr\hidewidth$\displaystyle\int$\hidewidth\cr}}
  {\ooalign{\raisebox{.14\height}{\scalebox{.7}{$\textstyle\sum$}}\cr\hidewidth$\textstyle\int$\hidewidth\cr}}
  {\ooalign{\raisebox{.2\height}{\scalebox{.6}{$\scriptstyle\sum$}}\cr$\scriptstyle\int$\cr}}
  {\ooalign{\raisebox{.2\height}{\scalebox{.6}{$\scriptstyle\sum$}}\cr$\scriptstyle\int$\cr}}
}

\begin{document}


\title{Fluctuations of the order parameter in an $SU(N_c)$ effective model}


\author{Pok Man \surname{Lo}}
\email[]{pokman.lo@uwr.edu.pl}
\affiliation{Institute of Theoretical Physics, University of Wroclaw,
PL-50204 Wroc\l aw, Poland}
\author{Krzysztof Redlich}
\affiliation{Institute of Theoretical Physics, University of Wroclaw,
PL-50204 Wroc\l aw, Poland}
\author{Chihiro Sasaki}
\affiliation{Institute of Theoretical Physics, University of Wroclaw,
PL-50204 Wroc\l aw, Poland}


\date{\today}

\begin{abstract}
We investigate features of the deconfinement phase transition in an $SU(N_c)$ gauge theory as revealed by fluctuations of the order parameter.  The tool of choice is an effective model built from one-loop expressions of the field determinants of gluon and ghost, in the presence of a Polyakov loop background field.  We show that the curvature masses associated with the Cartan angles, which serve as a proxy to study the $A_0$-gluon screening mass, show a characteristic dip in the vicinity of the transition temperature.  The strength of the observables, which reflects a competition between the confining and the deconfining forces, is sensitive to assumptions of dynamics, and thus provides an interesting link between the $Z(N_c)$ vacuum structure and the properties of gluon and ghost propagators.
\end{abstract}


\maketitle

\section{Introduction}
In this work we study the fluctuations of the order parameter in an $SU(N_c)$ gauge theory within an effective model.
Unlike the order parameter, these observables are finite and temperature
dependent even in the confined phase, thus providing important diagnostic
information about the mechanism of deconfinement phase transition, 
and the properties of gluons (and ghosts) in relation to the structure of $Z(N_c)$ vacuum.

Even when powerful numerical methods such as lattice QCD (LQCD) are available to
perform {\it ab initio} calculations of the full
theory~\cite{Boyd:1996bx,Borsanyi:2012ve,Kaczmarek:2002mc}, 
it is instructive, and sometimes essential, 
to work on an effective model description of a dynamical system.
First of all, it provides clear links between the observables and the underlying symmetry.
Second, it enables straightforward application of the model to
other extreme conditions~\cite{Fukushima:2013rx,Fukushima:2017csk}, or as a building block to study further coupling to
other dynamical fields~\cite{Andersen:2014xxa,Bruckmann:2013oba,Fraga:2013ova,Pagura:2016pwr,Lo:2018wdo,Lo:2020ptj}.

A common strategy to constructing an effective potential is via 
a polynomial of the order parameter
field~\cite{Ratti:2005jh,mat_model_1,Lo:2013hla}, 
i.e., the Ginzburg-Landau theory.
Symmetry restricts the kind of terms that can appear in the potential. 
The coefficients are generally smooth functions of temperatures ( and other external
fields ), which need to be separately determined, e.g. by fitting observables to
LQCD results. 

While a polynomial type potential is convenient to work with, 
the relation between model parameters and 
the properties of the underlying gluons (and ghosts) is not transparent.
In this study, we employ an effective potential built from one-loop expressions of the 
field determinants of gluon and ghost described in Ref.~\cite{Reinosa:2014ooa}.
(See also Ref.\cite{Braun:2007bx}.)
The model naturally describes both the confined and the deconfined phases, 
as related to the spontaneous breaking of $Z(N_c)$ symmetry.
In particular, the ghost term gives a confining, i.e., $Z(N_c)$ restoring, potential.
The effective model, as a tool, allows us to gain insights into the interplay
between vacuum structure and dynamics. 

The thermal properties of a pure gauge system have been analyzed previously 
by effective
models~\cite{Meisinger:2001cq,Meisinger:2003id,mat_model_1,gen_1,Bannur:2006js,Braun:2007bx,referee4,sasaki_pot,Fukushima:2012qa,Lo:2013hla}. 
However, features of gluons in the confined phase are usually not examined, 
and the importance of fluctuation observables~\cite{Lo:2013etb} has not been fully realized.
We therefore focus on these observables in this work and study how features of
deconfinement manifest through them.
We also use this opportunity to clarify the connection of these observables to
eigenvalues of the Polyakov loop operator in a matrix
model~\cite{mat_model_1,Meisinger:2001cq,Meisinger:2003id}.
We show that the curvature masses associated with the Cartan angles, 
which serve as a proxy to study the $A_0$-gluon screening mass, 
shows a characteristic trend of a rapid drop in the vicinity of transition
temperature $T_d$. Such a behavior is traceable to the competing effect of
$Z(3)$ restoring (confining) and $Z(3)$ breaking
(deconfining) forces. 
The strength of the masses is sensitive to the assumptions made on the dynamical
properties of gluon and ghost propagators.
Finally we present a possible relation between the glueball mass and $T_d$ suggested by the model.

\section{group structure of $SU(N_c)$}

The Polyakov loop operator in the fundamental representation, after a diagonalizing unitary transformation, 
can be expressed by the $N_c$ eigenphases $\vec{q}$:

\begin{equation}
    \hat{\ell}_F = {\rm diag} \, \left( e^{i q_1}, e^{i q_2}, \ldots, e^{i
    q_{Nc}} \right).
\end{equation}
The first $N_c-1$ phases may be taken as independent, 
and unitarity is enforced by requiring

\begin{equation}
    q_{N_c} = - \sum_{j=1}^{N_c-1} \, q_j.
\end{equation}
Alternatively, the angles can be expressed in terms of the
$N_c-1$ group angles of the maximal $Z(N_c)$ Cartan subgroup ($\gamma_j$'s),

\begin{equation}
    \vec{q} = \sum_{j=1}^{N_c-1} \, \gamma_j \, \vec{v}_j,
\end{equation}
where $\{ \vec{v}_j \} $ is a set of basis vectors, 
each being an $N_c$ dimensional vector with its sum of elements zero.
The order parameter field is obtained from a trace of $\hat{\ell}_F$,

\begin{equation}
    \ell = \frac{1}{N_c} \, {\rm Tr} \, \hat{\ell}_F.
\end{equation}
For $N_c \geq 3$, the order parameter is complex, and one can explore its real
and imaginary parts:

\begin{equation}
    \label{eq:xy}
    \begin{split}
    \ell &= X + i \, Y \\
    X &= \frac{1}{N_c} \, \sum_{j=1}^{N_c} \, \cos(q_j) \\
    Y &= \frac{1}{N_c} \, \sum_{j=1}^{N_c} \, \sin(q_j).
    \end{split}
\end{equation}
Note that $X, Y$ are regarded as a scalar function of the $N_c-1$ Cartan angles $\vec{\gamma}$'s.

To study the fluctuations of the order parameter in an effective model, 
we need to perform $X, Y$-field derivatives of a potential. 
Equation \ref{eq:xy} provides a connection between these derivatives with those acting on $\gamma_j$'s,

\begin{equation}
    \label{eq:dxy_op}
    \begin{split}
    \frac{d}{d X} &= \sum_{j=1}^{N_c-1} \, C_{1j}(\vec{\gamma}) \, \frac{d}{d
    \gamma_j} \\
    \frac{d}{d Y} &= \sum_{j=1}^{N_c-1} \, C_{2j}(\vec{\gamma}) \, \frac{d}{d
    \gamma_j},
    \end{split}
\end{equation}
where the $2 \times (N_c-1)$ matrix $C$ is obtained by (left) inverting the
transpose of the Jacobian $J$:

\begin{equation}
    \begin{split}
    J &= \frac{\partial \, \{X, Y\}}{\partial \, \{\gamma_1, \gamma_2, \ldots
    \gamma_{N_c-1} \}} \\
    C &= [J^t]^{-1}.
    \end{split}
\end{equation}

Finally, starting with a potential expressed in terms of the Cartan angles, 
$U(\vec{\gamma})$, the susceptibilities can be computed by forming
the curvature matrix $\bar{U}^{(2)}$~\cite{Sasaki:2006ww,Lo:2014vba,Lo:2018wdo}

\begin{equation}
    \label{eq:xycurva}
    \bar{U}^{(2)} = \frac{1}{T^4} \, 
    \begin{pmatrix}
        \frac{\partial^2 \, U}{\partial X \, \partial X}  &
        \frac{\partial^2 \, U}{\partial X \, \partial Y} \\
        \frac{\partial^2 \, U}{\partial Y \, \partial X} &
        \frac{\partial^2 \, U}{\partial Y \, \partial Y}
    \end{pmatrix}.
\end{equation}
The various $(X, Y)-$field derivatives are calculated according to Eq.~\eqref{eq:dxy_op}. 
Inverting the curvature matrix gives

\begin{equation}
    T^3 \, \tilde{\chi} = \left( \bar{U}^{(2)} \right)^{-1},
\end{equation}
with

\begin{equation}
    \begin{split}
        T^3 \, \chi_L &= T^3 \, \tilde{\chi}_{11} \\
        T^3 \, \chi_T &= T^3 \, \tilde{\chi}_{22}.
    \end{split}
\end{equation}
Note that the notions of longitudinal
and transverse directions~\cite{Lo:2013etb} correspond to real and imaginary
components along the real line,
but this is not so for other $Z(N_c)$ vacua.

To illustrate the computation of fluctuations, 
we consider a schematic effective potential (model A) of the form,

\begin{equation}
    \label{eq:potA}
    U = U_{\rm conf.} + U_{\rm glue},
\end{equation}
where the confining part is modeled by the group invariant measure
$H$~\cite{Fukushima:2003fw},

\begin{equation}
    \label{eq:pot1}
    U_{\rm conf.} = -\frac{b}{2} \, T \, \ln H.
\end{equation}
This potential is confining in the sense that it tends to drive the system
toward the $Z(N_c)$ symmetric vacuum ($\ell = 0$).
The deconfining part, which prefers the spontaneously broken $Z(N_c)$ vacuum,
is modeled as 
\begin{equation}
    \label{eq:pot2}
    \begin{split}
        U_{\rm glue} &= n_{\rm glue} \, T \, \int \frac{d^3 k}{(2 \pi)^3} \\
        &\times {\rm Tr}_A \, \ln ( I - \hat{\ell}_A \, e^{-\beta E_A(k)} ),
    \end{split}
\end{equation}
with $E_A(k) = \sqrt{k^2 + m_A^2}$.
In Sec.~\ref{sec4}, we shall investigate some alternative forms of the potential
and discuss issues of gauge dependence and inclusion of wave function
renormalizations.

As we are mainly interested in studying the influence from group structure, 
we shall keep the model parameters as simple as possible. 
In fact, we shall start with the parametrization:
$ b = \left( 0.1745 \, {\rm GeV} \right)^{3} $, $n_{\rm glue}=2$, and $m_A \approx 0.756 $ GeV. 
~\footnote{Such a value of gluon mass ($\approx 0.7$ GeV) is supported by calculations 
in different gauges. We have also checked that using the Gribov
dispersion relation $E(k) = \sqrt{k^2 + m_A^4/k^2}$~\cite{Zwanziger:2004np}
or imposing a UV cutoff for $m_A \rightarrow m_A \,
e^{-k^2/\Lambda^2}$~\cite{eric_gb}
does not lead to significant differences in the observables studied.}
Two group structures appear in this schematic model: the adjoint operator
$\hat{\ell}_A$
and the group invariant measure $H$. It is useful to express them in terms of
the eigenphases. For the former, 

\begin{equation}
    \label{eq:adj1}
    \hat{\ell}_A = {\rm diag} \, \left( e^{i Q_1}, e^{i Q_2}, \ldots, 
    e^{i Q_{{N_c}^2-1}} \right),
\end{equation}
with
\begin{equation}
    \label{eq:adj2}
    \vec{Q} = \left( 0, \ldots, 0; q_j-q_k, -(q_j-q_k) \right),
\end{equation}
for $j < k$, $j, k = 1, 2, \ldots N_c$. 
The adjoint angles are constructed from the root system~\cite{Georgi:1982jb,mat_model_1}, 
classified into Cartan and non-Cartan parts:
(a) $N_c-1$ zeros, representing the identity matrix element in $\hat{\ell}_A$;
(b) $N_c \, (N_c-1) / 2$ pairs of $q_i-q_j$'s for $i > j$ and terms with the opposite sign.

An intuitive way to understand the form of potential Eq.~\eqref{eq:pot2} is
to realize that the adjoint derivative operator for the gluon field, 
in the presence of a diagonal background field 
$\hat{q} = i \, \beta g \bar{A}_0 = \beta g \bar{A}_4$, reads

\begin{equation}
    \begin{split}
        \bar{D}^{\rm adj}_\mu \, \mathbf{M}  &= \partial_\mu \, \mathbf{M} +
        \delta_{\mu 0} \, \frac{1}{\beta} \, [
            \hat{q}, \mathbf{M} ].
    \end{split}
\end{equation}
The adjoint operator acts on an arbitrary $SU(N_c)$ matrix $\mathbf{M}$, 
and the latter has $N_c^2-1$ independent entries.
As $\hat{q}$ is diagonal, the $ij$th component of the commutator $[\hat{q},
\mathbf{M}]$ is given by~\cite{instanton,Dumitru:2013xna}

\begin{equation}
    (q_i - q_j) \, \mathbf{M}_{ij}.
\end{equation}
For $i \neq j$, the multiplying factors are exactly the
nontrivial entries of the adjoint angles $\vec{Q}$ in Eq.~\eqref{eq:adj2}. 
The remaining $N_c$ diagonal elements of $\mathbf{M}$, of which $N_c-1$ are
independent, are multiplied by $0$, i.e., the Cartan part of $\vec{Q}$. 
The effects of the background field is thus similar to introducing an imaginary chemical
potential for the $N_c^2-1$ independent components. 
In particular, the gauge field determinant can be constructed

\begin{equation}
    \label{eq:matsu}
    \begin{split}
        {\rm Tr} \, \ln \, \bar{D}^{2}_{\rm adj} &= \sum_a \, V \, \sumint \,
        \ln \, \left( (\omega_n + \frac{Q_a}{\beta})^2 + (\vec{k})^2 \right) \\
        &= 2 \, V \, \int \frac{d^3 k}{(2 \pi)^3} \, {\rm Tr}_A \, \ln \, ( I -
        \hat{\ell}_A \, e^{-\beta k} ) \\
        &+ (T=0).
    \end{split}
\end{equation}
where $\sumint$ denotes a Matsubara sum over the bosonic frequencies 
and an integral over momenta. 
From now on, we shall retain only the finite temperature piece.
Equation~\eqref{eq:pot2} is its simple extension to introducing a finite gluon mass.

Another group structure of interest is the invariant measure.
This can also be expressed in terms of the eigenphases $q_j$'s via

\begin{equation}
    \begin{split}
    H &= \prod_{j > k} \, \vert e^{i q_j}-e^{i q_k} \vert^2 \\
     &= \prod_{j > k} \, 4 \, \sin^2 \left( \frac{q_j-q_k}{2} \right).
    \end{split}
\end{equation}
Note that there are $  N_c \, (N_c-1) /2 $ pairs of $(j > k)$ in the product.
A fact that would prove useful later is the construction of the (logarithm of)
invariant measure from

\begin{equation}
    \label{eq:haar}
    \ln H = {\rm Tr}_A^\prime \, \ln \left( I - \hat{\ell}_A \right),
\end{equation}
where ${\rm Tr}^\prime$ denotes the partial trace over the non-Cartan roots 
to avoid irrelevant divergences from vanishing elements.
Hence the effective potential can be expressed as
\begin{equation}
    U_{\rm conf.} = -\frac{b}{2} \, T \, {\rm Tr}_A^\prime \,
    \ln \, (1 - {\hat{\ell}_A}),
\end{equation}
with $\hat{\ell}_A$ in Eqs.~\eqref{eq:adj1} and~\eqref{eq:adj2}. 
This establishes that an invariant measure term behaves as the glue
potential~\eqref{eq:pot2} with $E_A(k) \rightarrow 0$, 
but of the opposite sign, and should be formally understood as a ghost
contribution~\cite{Weiss:1980rj,Gocksch:1993iy}. 

Here we explicitly work out the case for $N_c = 2, 3, 4$.

\subsection{$N_c=2$}

In this case there is only a single independent eigenphase $\vec{q} = (q_1, -q_1)$ for the Polyakov loop
operator $\hat{\ell}_F$,

\begin{equation}
    \hat{\ell}_F = {\rm diag} \, \left( e^{i q_1}, e^{-i q_1} \right),
\end{equation}
and the order parameter field is purely real,
\begin{equation}
    \ell = \cos q_1.
\end{equation}
The adjoint angles can be constructed

\begin{equation}
\vec{Q} = (0; 2 q_1, -2 q_1),
\end{equation}
and from Eq.~\eqref{eq:haar} the invariant measure works out to be
\begin{equation}
    \label{eq:hsu2}
    \begin{split}
    H(q_1) &= 4 \, \sin^2 q_1 \\
           &= 4 \, (1-\ell^2).
    \end{split}
\end{equation}

The same result may be obtained from a slightly different starting point. 
Consider the parametrization of $SU(2)$ matrices $\{u\}$ by $(a_0, \vec{a})$ via

\begin{equation}
    u = a_0 \, I + i \, \vec{a} \cdot \vec{\sigma},
\end{equation}
where $I$ and $\vec{\sigma}$ are the $ 2 \times 2$ identity and Pauli matrices.
The invariant measure is given by

\begin{equation}
    \begin{split}
    \int d \mu_{\rm SU(2)} &= \int d^4 a \, \delta(a^2-1) \\
                           &= \int d a_0 \,  d \vert \vec{a} \vert \, d^3 \hat{n}
                           \, \vert \vec{a} \vert^2 \, \delta(a_0^2+\vec{a}^2-1) \\
                           &\propto \int d a_0 \, \sqrt{1-a_0^2}.
    \end{split}
\end{equation}
The last line assumes a uniform distribution of $d^3 \hat{n}$.
We see that $a_0$ plays the role of $\ell = \cos(q_1)$:
The change of coordinates from $\ell$ to $q_1$ gives an extra factor of the
square root term, leading to the same expression of the invariant measure in
Eq.~\eqref{eq:hsu2}.

\subsection{$N_c=3$}

For the SU(3) gauge group there are two independent eigenphases $\vec{q} = (q_1,
q_2, -q_1-q_2)$; hence

\begin{equation}
    \hat{\ell}_F = {\rm diag} \, \left(e^{i q_1}, e^{i q_2}, e^{-i (q_1+q_2)} \right),
\end{equation}
and the order parameter field is
\begin{equation}
    \begin{split}
    \ell &= X + i \, Y \\
       X &= \frac{1}{3} \, (\cos q_1 + \cos q_2 + \cos (q_1+q_2)) \\
       Y &= \frac{1}{3} \, (\sin q_1 + \sin q_2 - \sin (q_1+q_2)).
    \end{split}
\end{equation}

\begin{table}[h!]
    \centering
    \begin{tabular}{ |c|c|c| } 
\hline
    $Q_1$ & $Q_2$ &  \\
\hline
    $0$ & $0$ &  \\
\hline
\hline
    $Q_3 = -Q_6 $ & $Q_4 = -Q_7$ & $Q_5=-Q_8$  \\
\hline
    $q_1-q_2$ & $2q_1+q_2$ & $q_1+2q_2$  \\
\hline
\end{tabular}
    \caption{Adjoint angles of the Polyakov loop operator for the $SU(3)$ group.}
    \label{tab:su3q}
\end{table}

The adjoint angles are shown in Table~\ref{tab:su3q}. 
Using these with Eq.~\eqref{eq:haar} the invariant measure can be computed

\begin{equation}
    \label{eq:hsu3}
    \begin{split}
        H(q_1, q_2) &= 64 \, \sin^2 \frac{(q_1-q_2)}{2}  \\
        &\times \sin^2 \frac{(2 q_1 + q_2)}{2} \, \sin^2 \frac{(q_1 + 2 q_2)}{2}.
    \end{split}
\end{equation}
We can also express the result in terms of the Cartan parameters. 
The two independent directions can be chosen to be

\begin{equation}
    \begin{split}
    \vec{v}_1 &= (1, 0, -1), \\
    \vec{v}_2 &= (1/2, -1, 1/2);
    \end{split}
\end{equation}

\begin{equation}
    \label{eq:su3gamma}
    \begin{split}
    \gamma_1 &= q_1 + {q_2}/2, \\
    \gamma_2 &= -q_2
    \end{split}
\end{equation}
can be taken as independent variables; and the invariant measure reads

\begin{equation}
    \begin{split}
        H(\gamma_1, \gamma_2) &\propto \sin^2 \frac{(\gamma_1 - 3/2 \,
        \gamma_2)}{2} \\
        &\times \sin^2 \gamma_1 \, \sin^2 \frac{(\gamma_1 + 3/2 \, \gamma_2)}{2}.
    \end{split}
\end{equation}

Specific to the SU(3) gauge group, the two independent degrees of freedom can
be identified with the trace of the Polyakov loop operator 
in the fundamental representation $\ell$, and the invariant
measure can be expressed via $X, Y$:

\begin{equation}
    \begin{split}
    H(X, Y) &= 27 \times ( \,  1 - 6 ( X^2 + Y^2 ) \\
    &+ 8 (X^3 - 3 X Y^2) - 3 (X^2+Y^2)^2
    \, ).
    \end{split}
\end{equation}
For $N_c > 3$, the invariant measure generally depends on $N_c-1$ independent angles, 
and therefore is not expressible solely in terms of $(X, Y)$.

\begin{figure*}[!ht]
\centering
\includegraphics[width=0.49\linewidth]{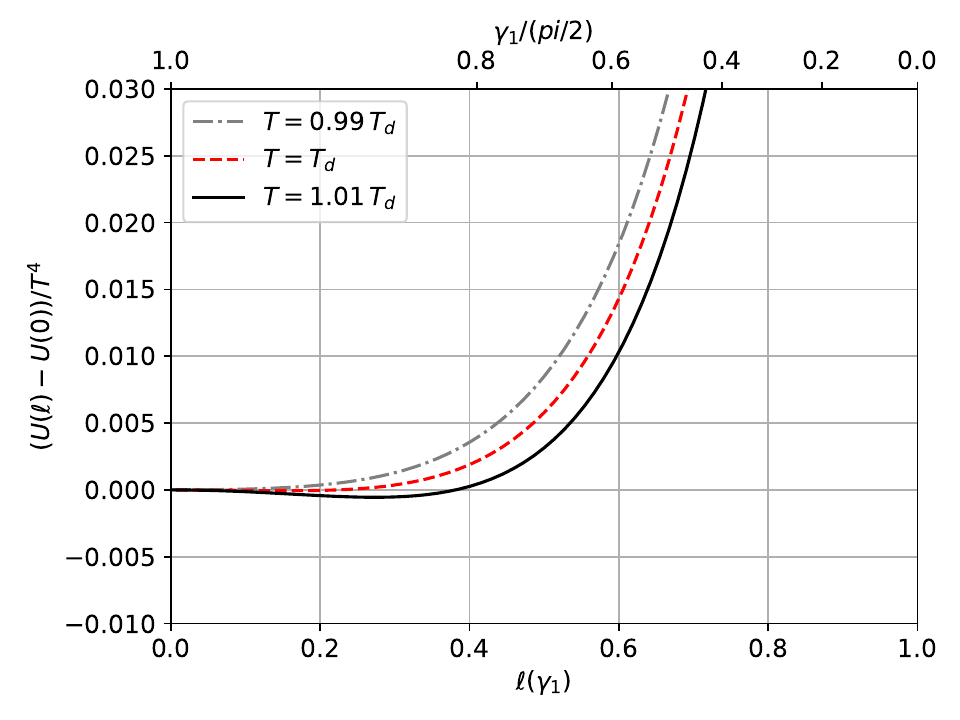}
\includegraphics[width=0.49\linewidth]{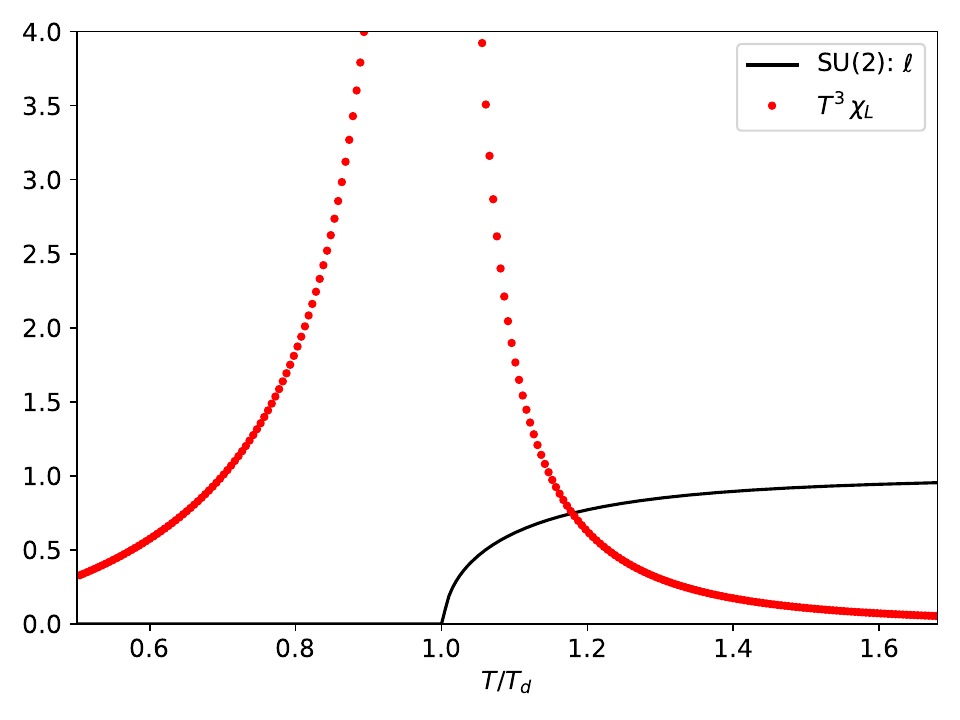}
\includegraphics[width=0.49\linewidth]{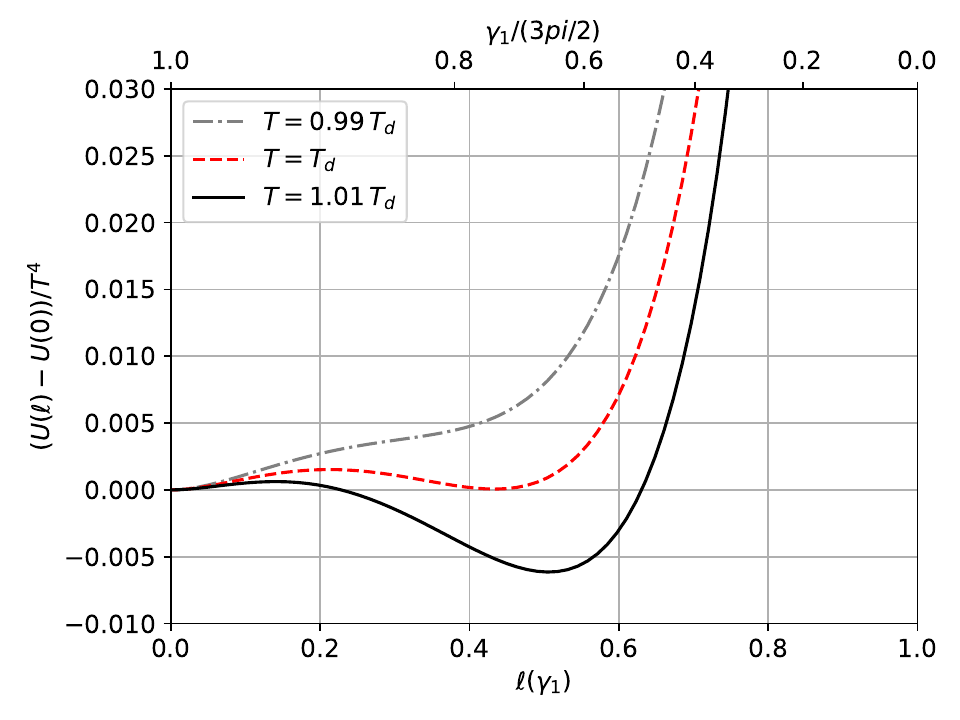}
\includegraphics[width=0.49\linewidth]{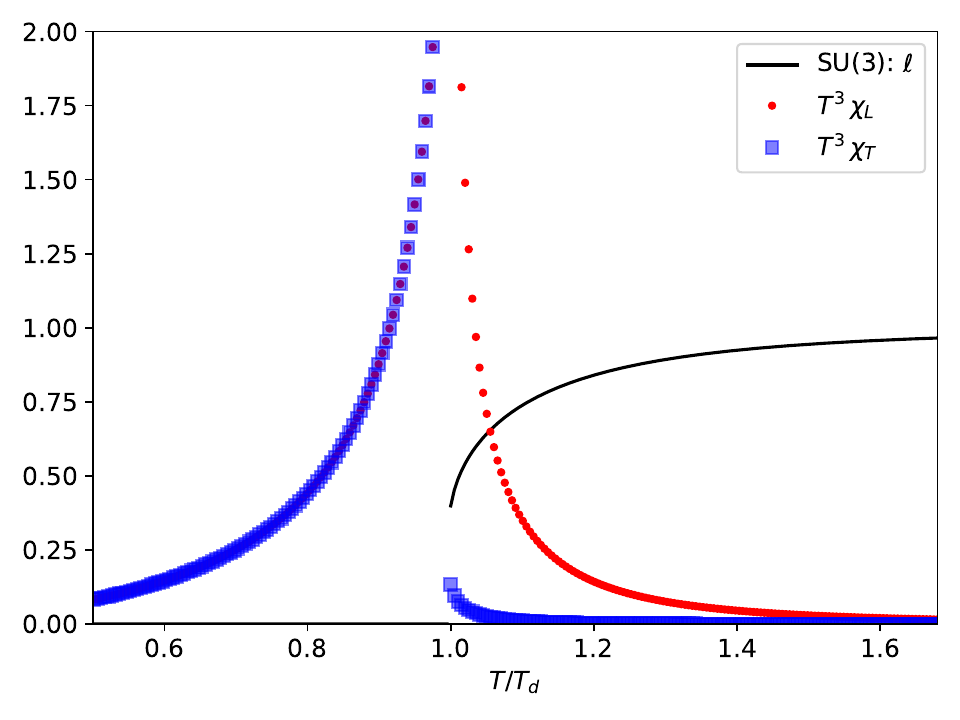}
\includegraphics[width=0.49\linewidth]{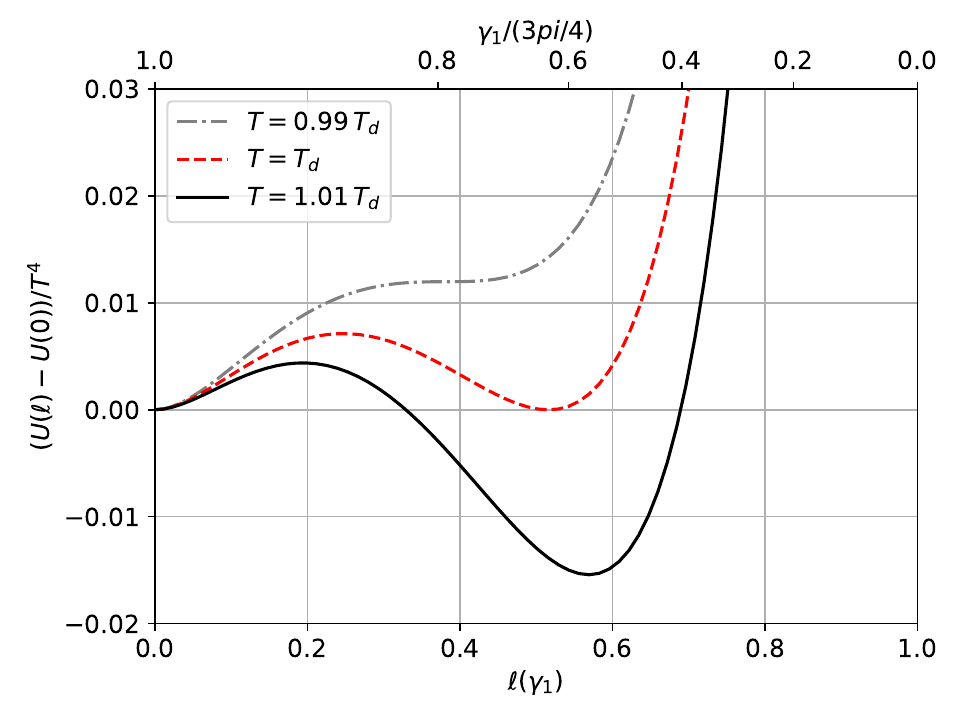}
\includegraphics[width=0.49\linewidth]{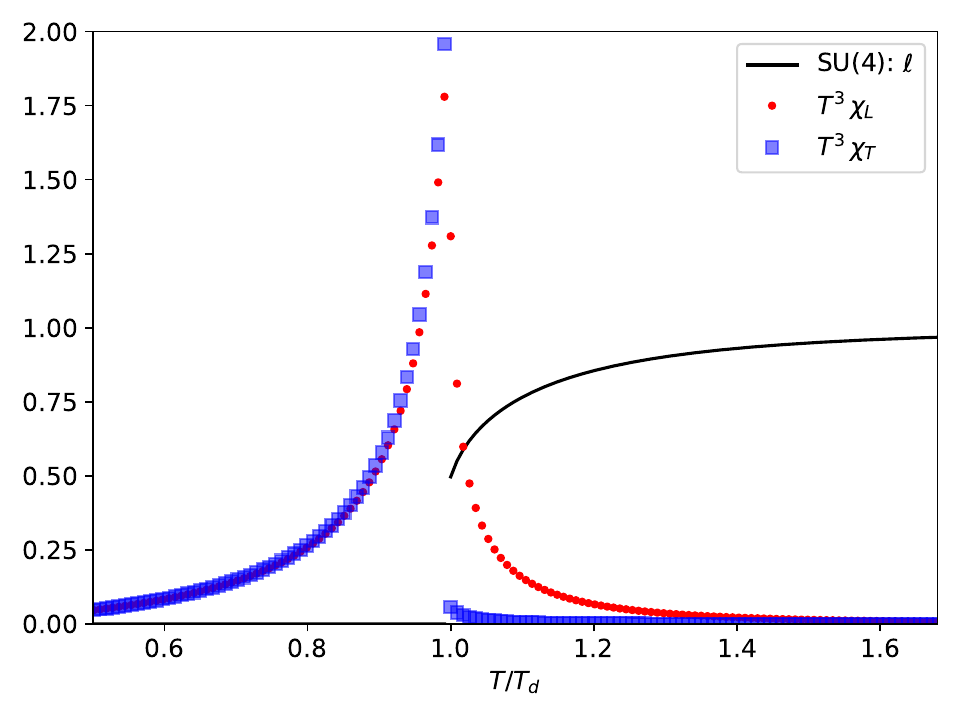}
        \caption{The Polyakov loop potentials~\eqref{eq:potA} (left) 
        and the derived observables: the Polyakov loop expectation values, the longitudinal and
        the transverse susceptibilities (right) for $N_c=2, 3, 4$.}
\label{fig1}
\end{figure*}

\subsection{$N_c=4$}

The analysis for $N_c=4$ and beyond proceeds in a similar fashion.
For the SU(4) gauge group there are three independent eigenphases:

\begin{equation}
    \vec{q} = (q_1, q_2, q_3, q_4)
\end{equation}
with $ q_4 = -(q_1 + q_2 + q_3)$. 

The order parameter field is given by
\begin{equation}
    \begin{split}
    \ell &= X + i \, Y \\
       X &= \frac{1}{4} \, (\cos q_1 + \cos q_2 + \cos q_3 + \cos (q_1+q_2+q_3)) \\
       Y &= \frac{1}{4} \, (\sin q_1 + \sin q_2 + \sin q_3 - \sin (q_1+q_2+q_3)).
    \end{split}
\end{equation}
The $15=3+6+6$ adjoint angles are composed of three zeros, six nontrivial
angles, and their negative values. See Table~\ref{tab:su4q}. 

\begin{table}[h!]
    \centering
    \begin{tabular}{ |c|c|c| } 
\hline
        $Q_1$ & $Q_2$ & $Q_3$ \\
\hline
        $0$ & $0$ & $0$  \\
\hline
\hline
        $Q_4=-Q_{10}$ & $Q_5=-Q_{11}$ & $Q_6=-Q_{12}$  \\
\hline
        $q_1-q_2$ & $q_1-q_3$ & $q_2-q_3$ \\
\hline
\hline
        $Q_7=-Q_{13}$ & $Q_8=-Q_{14}$ & $Q_9=-Q_{15}$  \\
\hline
        $2q_1+q_2+q_3$ & $q_1+2q_2+q_3$ & $q_1+q_2+2q_3$ \\
\hline
\end{tabular}
    \caption{Adjoint angles of the Polyakov loop operator for the $SU(4)$ group.}
    \label{tab:su4q}
\end{table}

The invariant measure can be constructed from the nontrivial adjoint angles:
\begin{equation}
    H(\vec{q}) \propto \prod_{j=4-9} \, \sin^2 Q_j.
\end{equation}
To translate this into the Cartan $\vec{\gamma}$, we can use the following basis vectors:

\begin{equation}
    \label{eq:su4}
    \begin{split}
    \vec{v}_1 &= (1, 1/3, -1/3, -1), \\
    \vec{v}_2 &= (1, -1, -1, 1), \\
    \vec{v}_3 &= (1/3, -1, 1, -1/3).
    \end{split}
\end{equation}
In particular, going along $\vec{v}_1$, corresponding to the uniform eigenvalue
ansatz~\cite{mat_model_1}, the order parameter field is purely real,
and through $\gamma_1$ we can relate the invariant measure to the Polyakov loop:

\begin{equation}
    \begin{split}
       X &= \frac{1}{2} \, (\cos \gamma_1 + \cos (\gamma_1/3) ), \\
       Y &= 0,
    \end{split}
\end{equation}
and

\begin{equation}
    H(\vec{q} \rightarrow \gamma_1 \vec{v}_1) \propto \sin^6 (\gamma_1/3) \, \sin^4 (2 \gamma_1/3)  \, \sin^2 (\gamma_1).
\end{equation}
compared to a similar projection in the $SU(3)$ case

\begin{equation}
    H(\vec{q} \rightarrow \gamma_1 \vec{v}_1)) \propto \sin^2 \frac{\gamma_1}{2}
    \,
    \sin^2 \gamma_1 \,
    \sin^2 \frac{\gamma_1}{2}.
\end{equation}

\section{Polyakov loop and the susceptibilities}

\subsection{General results}
Once an effective potential is specified, 
its minimization and the extraction of various observables 
are standard procedure~\cite{Lo:2013hla}. 
Here we simply display the results in Fig.~\ref{fig1} and highlight some observations:

\begin{itemize}

    \item First, the order of the phase transition naturally changes from second
        order for $N_c=2$ to first order for $N_c \geq 3$.
        Note that the same set of model parameters has been used in the
        calculations.

    \item Second, the two susceptibilities derived for $N_c \geq 3$ are equal in
        the confined phase, and a narrow aspect ratio for the shape of the potential, 
        i.e., $\chi_T \ll \chi_L$ in the deconfined phase. 
        This case is known for
        $N_c=3$~\cite{Lo:2013etb}. 
        Equation~\eqref{eq:dxy_op} makes it possible to study the fluctuations
        beyond $N_c=3$, and for the first time we can verify a similar trend is
        observed in this class of model for $N_c \geq 4$ under the uniform
        eigenvalue ansatz~\cite{mat_model_1}.

    \item It is expected that the first order phase transition becomes stronger
        as $N_c$ increases. This is the case in this model, 
        and comparing the $N_c=4$ case with $N_c=3$, 
        we observe the Polyakov loop at $T_d$ increases, 
        while the magnitudes of the susceptibilities decrease. 
        The latter suggests larger curvatures of the potential around the minima,
        which sets the stage for a stronger phase transition.
        As $N_c$ increases further, we find that $\ell(T_d)$ tends to $\approx
        0.5$, while the decreasing trend of the susceptibilities
        continues.~\footnote{The value becomes $\ell(T_d, N_c\rightarrow \infty) \approx 0.6$ for models B and C introduced
            later. See Eq.~\eqref{eq:modelB}.}

\end{itemize}

\begin{figure}[!ht]
\centering
\includegraphics[width=\linewidth]{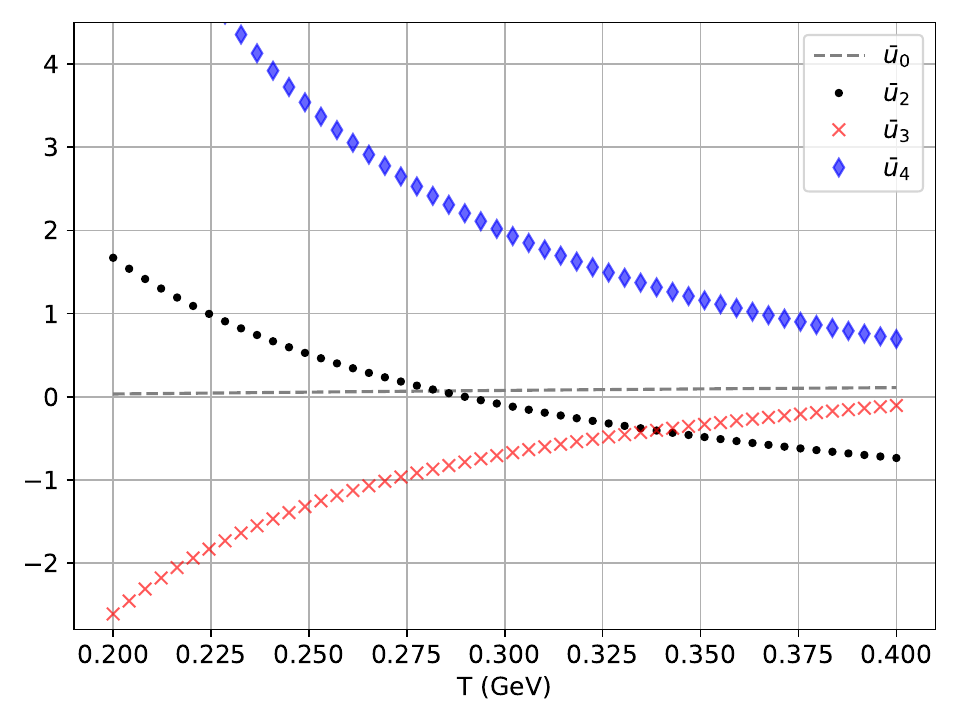}
        \caption{Landau parameters Eq.~\eqref{eq:landau} derived from the effective
        potential~\eqref{eq:potA} as functions of temperature.}
\label{fig2}
\end{figure}

The Landau parameters can be directly extracted in this model. 
For the case of $SU(3)$ along the real line, we write 

\begin{equation}
    \frac{U}{T^4} = \bar{u}_0 + \bar{u}_2 \, X^2 + \bar{u}_3 \, X^3 + \bar{u}_4 \, X^4 + \cdots. 
\end{equation}
Expanding potentials~\eqref{eq:pot1} and ~\eqref{eq:pot2} in powers of
$X$, we obtain (in the Boltzmann limit)

\begin{equation}
    \label{eq:landau}
    \begin{split}
        \bar{u}_0 &= \frac{1}{\pi^2} \,
        ( \frac{m_A}{T} )^2 \, K_2(\frac{m_A}{T}), \\
        \bar{u}_2 &= \frac{3 b}{T^3} - \frac{9}{\pi^2} \,
        ( \frac{m_A}{T} )^2 \, K_2(\frac{m_A}{T}), \\
        \bar{u}_3 &= - \frac{4 b}{T^3} + \frac{27}{\pi^2} \,
        (\frac{m_A}{T} )^2 \, K_2(\frac{2 m_A}{T}), \\
        \bar{u}_4 &= \frac{21 b}{2 T^3} - \frac{81}{4 \pi^2} \,
        (\frac{m_A}{T} )^2 \, K_2(\frac{2 m_A}{T}),
    \end{split}
\end{equation}
where $K_2$ is the modified Bessel function of the second kind (order $2$).
These relations link the Landau parameters to properties of the
underlying gluons.
To derive these results we have used the fact that

\begin{equation}
    \begin{split}
        {\rm Tr} \, \hat{\ell}_A &=  \left( {\rm Tr} \, \hat{\ell}_F \right)^2 - 1 \\
        &\xrightarrow{SU(3)} 9 \, X^2 -1, \\
        {\rm Tr} \, \hat{\ell}_A^2 &=  \left( {\rm Tr} \, \hat{\ell}_F^2 \right)^2 - 1 \\
        &\xrightarrow{SU(3)} 36 \, X^2 - 108 \, X^3 + 81 \, X^4 - 1
    \end{split}
\end{equation}
along the real line. 
The expansion works best in the confined phase, where $X \ll 1$.
Note that the cubic term arises naturally from ${\rm Tr} \, \hat{\ell}_A^2$,
and we can readily verify the standard scenario of a first
order phase transition: $\bar{u}_3 < 0, \bar{u}_4 > 0$, and $\bar{u}_2$ changes sign (from positive to
negative) close to $T_d$. See Fig.~\ref{fig2}.
The susceptibilities can be simply constructed from

\begin{equation}
    (T^3 \, \chi_{L,T})^{-1} \approx (2 \bar{u}_2).
\end{equation}
Thus, the observables are driven by a competition between the confining and
the deconfining potentials. 
This is a general observation for the class of models we study.
Also the condition 

\begin{equation}
    \bar{u}_2(T) = 0
\end{equation}
is useful for a qualitative understanding of $T_d$,
giving $T \approx 0.29 $ GeV instead of the
true value  $ T_d = 0.274 $ GeV, calculated numerically.

\subsection{Gluon density in the presence of Polyakov loop}

\begin{figure}[!ht]
\centering
\includegraphics[width=\linewidth]{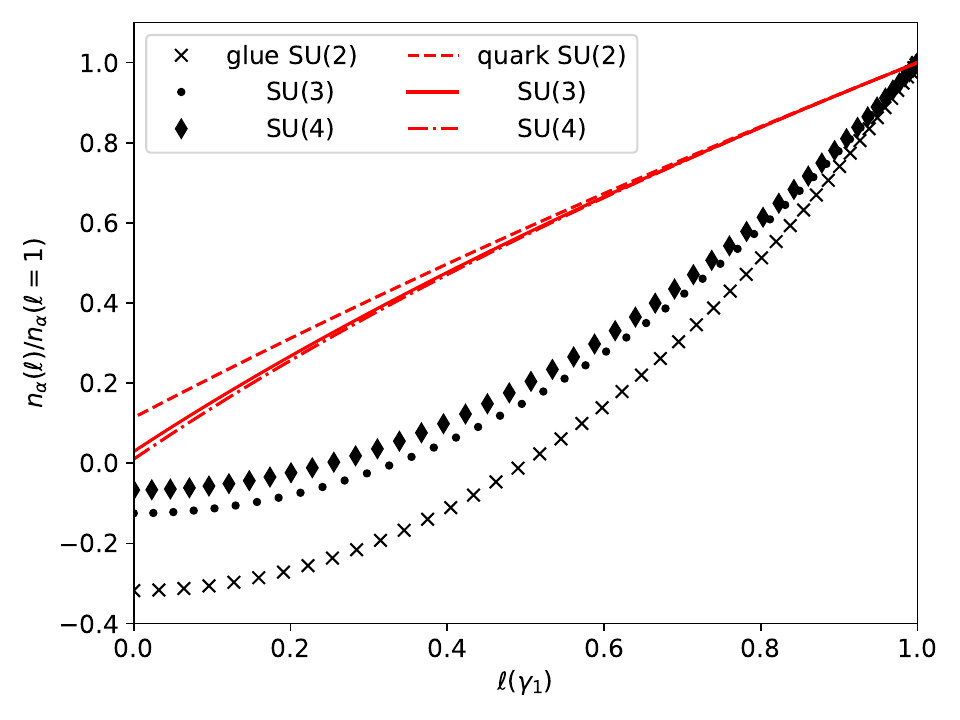}
        \caption{Thermal densities of quarks and gluons in the presence of a
        background Polyakov loop field for $N_c=2, 3, 4$. The results are
        evaluated at $T = 0.24$ GeV with an effective gluon mass $m_A = 0.756$
        GeV and a quark mass $m_Q = 0.1$ GeV.}
\label{fig3}
\end{figure}

A key feature of an effective Polyakov loop model is a description of the thermal densities of gluons and quarks in the presence of a Polyakov loop mean field.
These can be conveniently expressed in terms of the eigenphases. 
Take, for example, the case for SU(3), along the real line, they depend only on a
single angle variable $\gamma_1$ (i.e., $\gamma_2 = 0$),

\begin{equation}
    \hat{\ell}_F \rightarrow {\rm diag} \, \left( e^{i \gamma_1}, 1, e^{-i
    \gamma_1} \right)
\end{equation}
and

\begin{equation}
    \begin{split}
        \hat{\ell}_A \rightarrow {\rm diag} \, &( 1, 1, e^{i \gamma_1}, e^{2 i \gamma_1}, e^{i \gamma_1}, \\
        & e^{-i \gamma_1}, e^{-2 i \gamma_1}, e^{-i \gamma_1} ),
    \end{split}
\end{equation}
such that

\begin{equation}
    \label{eq:glue_den}
    n_{\rm glue}(\gamma_1) = \int \frac{d^3 k}{(2 \pi)^3} \, \sum_{i=1-8}
    \frac{\hat{\ell}_A^{(j)}}{e^{\beta E_A}-\hat{\ell}_A^{(j)}}.
\end{equation}
Note that as $\gamma_1 \rightarrow 0$, $\hat{\ell}_A$ becomes an identity in the
$8 \times 8$ adjoint space, and Eq.~\eqref{eq:glue_den} recovers the free
quantum Bose gas limit, 

\begin{equation}
    n_{\rm glue}(\gamma_1 \rightarrow 0) = 8 \times \int \frac{d^3 k}{(2
    \pi)^3} \, \frac{1}{e^{\beta E_A(k)}-1}.
\end{equation}

An analogous expression can be derived for quarks, 
except that the trace is over the entries 
of the Polyakov loop operator in the fundamental representation $\hat{\ell}_F$.
It can also be expressed as a function of $\gamma_1$:

\begin{equation}
    n_{\rm quarks}(\gamma_1) = \int \frac{d^3 k}{(2 \pi)^3} \, \sum_{i=1-3}
    \frac{\hat{\ell}_F^{(j)}}{e^{\beta E_Q(k)}+\hat{\ell}_F^{(j)}}.
\end{equation}
Similarly the free quantum fermion gas limit is recovered at $\gamma_1
\rightarrow 0$,
\begin{equation}
    n_{\rm quarks}(\gamma_1 \rightarrow 0) = 3 \times \int \frac{d^3 k}{(2 \pi)^3}
    \, \frac{1}{e^{\beta E_Q(k)}+1}.
\end{equation}

A plot of these thermal densities are shown in Fig.~\ref{fig3}, illustrated for
the case of $N_c = 2, 3, 4$. 
Note that the x axis is the corresponding traced Polyakov loops, projected along the real line:

\begin{equation}
    \begin{split}
    \ell_{\rm SU(2)}(\gamma_1) &= \cos \gamma_1, \\
    \ell_{\rm SU(3)}(\gamma_1) &= \frac{1}{3} \, \left( 1 + 2 \, \cos \gamma_1
        \right), \\
    \ell_{\rm SU(4)}(\gamma_1) &= \frac{1}{2} \, \left( \cos \gamma_1 + \cos
    (\gamma_1/3) \right).
    \end{split}
\end{equation}

An important observation is that both densities are substantially suppressed at $\ell \rightarrow 0$
compared to the free gas limits $\ell \rightarrow 1$. 
This is how confinement is represented in this class of models: a statistical
confinement that relates the thermal
abundances of quarks and gluons to the expectation value of the $Z(N_c)$ order
parameter, i.e., the Polyakov loop. 

For gluons, the $\ell = 0$ limit turns mildly negative.
This does not necessarily mean we have a negative pressure in the bulk, 
since other contributions, such as those coming from the confining ghosts,
can reverse this negative value.
The thermal distribution of confining gluons in the QCD medium is still an open issue, 
and a small negative value in some momentum range is not ruled out. 
Nevertheless, it is likely that the beyond one-loop, nonperturbative
interactions will produce further corrections to this quantity.~\footnote{The standard
prescription in an effective model is to simply subtract the potential at $\ell
= 0$. This fixes the problem of negative partial pressure and density of gluons
in the confined phase. Of course, results for the Polyakov loops and their fluctuations
are not affected.  However, one then needs to correct for the right number of
gluonic degrees of freedom in the deconfined phase at high temperatures.}
A similar plot from a nonperturbative study, such as Schwinger Dyson equations
in a given gauge, can help to clarify the
issue~\cite{vonSmekal:1997ern,vonSmekal:1997ohs,Fischer:2008uz,Aguilar:2008xm,
referee1,referee2,referee3,Maas:2011se}.

In any case the suppression discussed here,
linked to an order parameter for the spontaneous $Z(N_c)$ breaking, 
is an effective description.
There are interesting differences from models which predict the suppression in
the spectrum via an infrared divergent mass (and wave function) renormalizations~\cite{Lo:2009ud}. 
The task remains to understand the connections between various models of
confinement, and to further explore the dynamical aspects of confining
quasigluons in the QCD medium, 
practical for building phenomenology of the thermal system of glueballs and other
objects.

\section{curvature masses of Cartan angles}
\label{sec4}

\subsection{Gauge dependence and effects of wave function renormalization}

In Ref.~\cite{Reinosa:2014ooa} the phase transition of the pure Yang-Mills system is studied using
the background field method in the Landau-DeWitt gauge. 
A confining potential is motivated from the ghost determinant:

\begin{equation}
        {\rm Tr} \, \ln \, \bar{D}^{2}_{\rm adj}.
\end{equation}
This gives a potential of exactly the same form as $U_{\rm glue}$ in 
Eq.~\eqref{eq:pot2}, but with an opposite sign, and is hence $Z(N_c)$
restoring.
Note that the correct way to account for a ghost contribution 
to the thermodynamic potential involves a bosonic Matsubara sum of the relevant
operator (with a factor of $-1$), instead of a fermionic Matsubara sum.~\cite{Bernard}
The total potential can be written as

\begin{equation}
    U_{\rm tot} = 3 U_1(m_A) - U_1 (0).
\end{equation}
This form makes it obvious that we are considering three gluons and one ghost. 
Both terms can be expressed with $U_1$ that reads

\begin{equation}
    \begin{split}
    U_1(m_A) &= \frac{1}{2 \beta} \, \sum_a \, \sumint \, \ln \, \left(
        \tilde{k}_a^2 + m_A^2 \right) \\
        &= T \, \int \frac{d^3 k}{(2 \pi)^3} \, {\rm Tr}_A \, \ln \, ( I -
        \hat{\ell}_A \, e^{-\beta E_A(k)} )
    \end{split}
\end{equation}
with
\begin{equation}
    \label{eq:ktilde}
    \tilde{k}_a^2 = (\omega_n + \frac{Q_a}{\beta})^2 + (\vec{k})^2.
\end{equation}
Note that the invariant measure term~\eqref{eq:haar} can be regarded as a limiting case of
$U_1$ with $E_A \rightarrow 0$.~\footnote{In the axial
gauge~\cite{Weiss:1980rj}, it was argued the invariant measure term is 
canceled by a similar term in the glue potential. The case of massive gluons remains to be explored.}
One can speculate the form of potential in the 't Hooft-Feynman gauge to read

\begin{equation}
    \begin{split}
        U_{\rm tot} &= 2 U_1 (m_A) + \Delta U_\xi \\
        \Delta U_\xi &= (1+\Delta n_\xi) \times ( U_1(m_A) - U_1 (0) ).
    \end{split}
\end{equation}
The subscript $\xi$ signifies the possible gauge dependence, 
e.g., $\Delta n_\xi = 0 \, (1)$ in Landau-DeWitt ('t Hooft-Feynman) gauge.
Note that the gluon mass $m_A$ itself can depend on the gauge choice.
Nevertheless, in all cases the physical limit of 2 gluonic degrees of freedom
at high temperature (the Stefan-Boltzmann limit) 
is recovered when we set $m_A=0$, $\hat{\ell}_A = I$ in the perturbative vacuum.

\begin{figure}[!ht]
\includegraphics[width=\linewidth]{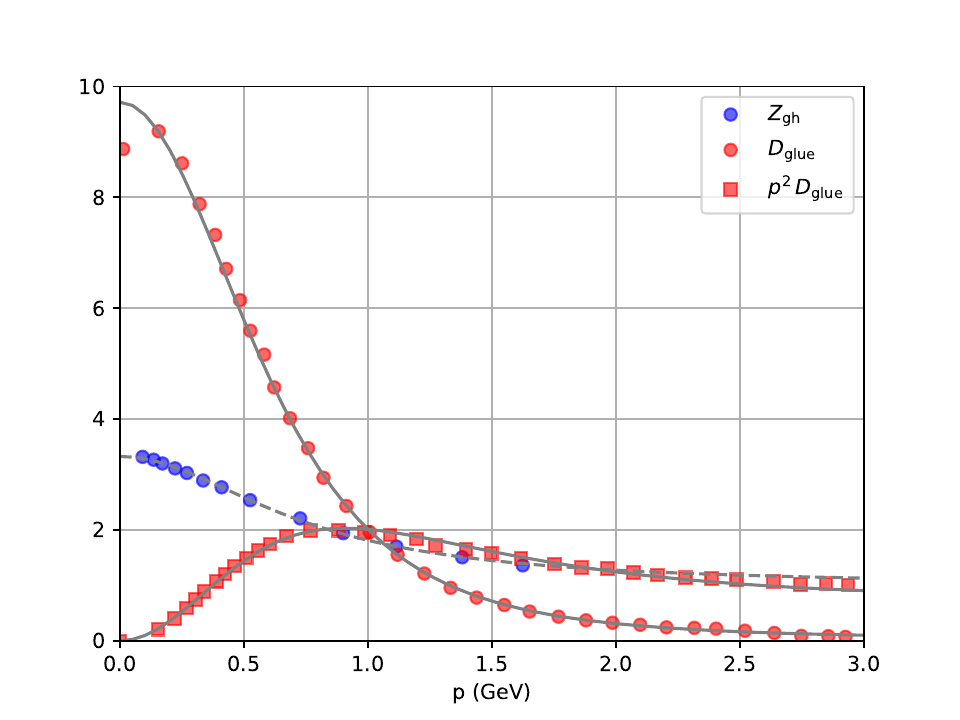}
        \caption{Fits of ($T=0$) LQCD results~\cite{Bogolubsky:2009dc} of gluon propagator and ghost
        wave function renormalization 
        with the generalized Gribov-Stingl form in Eqs.~\eqref{eq:gh1} and
        ~\eqref{eq:glue1}.}
\label{fig4}
\end{figure}

We next expand the model to include effects of wave function renormalizations of gluons and ghosts~\cite{Fukushima:2012qa}.
Assuming the background field continues to enter as Eq.~\eqref{eq:ktilde},
e.g., when the ghost propagator is modified by

\begin{equation}
    \frac{1}{\tilde{k}^2} \rightarrow \frac{Z_{\rm
    gh}(\tilde{k}^2)}{\tilde{k}^2},
\end{equation}
the corresponding change in the potential reads

\begin{equation}
    {\rm Tr} \, \ln \, \tilde{k}^2 \rightarrow {\rm Tr} \, \left( \ln
    \, \tilde{k}^2 - \ln \, Z_{\rm gh} (\tilde{k}^2) \right).
\end{equation}
A further simplification is possible if we approximate $Z_{\rm gh}$ in the 
Gribov-Stingl form~\cite{Fukushima:2012qa}:

\begin{equation}
    \label{eq:gh1}
    Z_{\rm gh} \propto (\frac{\tilde{k}^2 + R_1^2}{\tilde{k}^2 +
    R_2^2})
\end{equation}
with some mass scales $R_{1,2}$. 
Note that $Z_{\rm gh}(\tilde{k}^2 \rightarrow \infty) \rightarrow 1$.
The corresponding modification in the effective Polyakov loop potential is
given by

\begin{equation}
    \label{eq:gh2}
    U_1(0) \rightarrow U_1(0) - ( U_1(R_1) -  U_1(R_2) )
\end{equation}
for each ghost field. 
For demonstration, 
we fit the result of the lattice determination 
of the wave function renormalization of the ghost
propagator~\cite{Bogolubsky:2009dc} (in Landau gauge) with the
parametrization~\eqref{eq:gh1}. 
A reasonable fit is obtained with parameters $R_1 = 1.335$ GeV and $R_2=0.732$
GeV. A similar scheme can be applied to the gluons, with a slightly modified
form:

\begin{equation}
    \label{eq:glue1}
    \begin{split}
    Z_{\rm glue} &= (\frac{\tilde{k}^2 + R_1^2}{\tilde{k}^2 + 
        R_2^2})^{g_1} \, (\frac{\tilde{k}^2 + R_3^2}{\tilde{k}^2 +
        R_4^2})^{g_2} \\
        D_{\rm glue} &= \frac{Z_{\rm glue}}{\tilde{k}^2 + m_A^2}.
    \end{split}
\end{equation}
The parameters are $(g_1, R_1, R_2) = (4, 2.615, 1.660)$, $(g_2, R_3, R_4) = (1, 2.616, 6.794)$,
and $m_A = 0.756$, all in appropriate units of GeVs. 
The results are shown in Fig.~\ref{fig4}.

The change in the effective potential can be intuitively understood as follows: 
The enhancement of $Z_{\rm gh}$ at low momenta dictates $R_1 > R_2$, 
with a stronger Boltzmann suppression gives $\vert U_1(R_1) \vert < \vert
U_1(R_2) \vert$, 
and finally leads to a strengthening of the ghost potential (while preserving its sign). 
See Eq.~\eqref{eq:gh2}. 
~\footnote{It is also possible that the function drops rapidly to zero in the
deep infrared, and hence the form~\eqref{eq:gh2} needs to be
modified~\cite{Alkofer:2000wg,Fischer:2008uz,Iritani:2009mp,Maas:2017csm,Maas:2019ggf}.
Furthermore, there are more refined studies on decomposing the gluon
propagators into $Z_{\rm glue}(k)$ and mass function $m_A(k)$~\cite{orlando1,orlando2}.
The scenario in other gauges, 
and the efficacy of the commonly used approximation schemes, 
such as static approximation or expansions around simple poles, 
should be investigated in the future.}
A stronger potential is also found when implementing the
lattice result of the gluon propagator with a wave function renormalization. 
On the other hand, the value of $T_d$ depends on the competition between the two 
and requires an actual calculation to deduce the trend.

We thus obtain a unified framework to discuss the modeling of an effective
potential in different approximation schemes:

\begin{equation}
U_{\rm tot} = 2 U_1 (m_A) + \Delta U_\xi
\end{equation}
with

\begin{equation}
    \begin{split}
        \Delta U_\xi &= (1 + \Delta n_\xi) \, ( U_1(m_A) -  U_1(0) ) \\
    &+ \sum_j \,
    g_j \, ( U_1(R_1^{(j)}) -  U_1(R_2^{(j)}) ).
    \end{split}
\end{equation}
The key observation is that the same group structure appears in
various contributions to the potential, 
and details of gluon and ghost propagators are subsumed into the model parameters. 
The effective framework thus provides a transparent way 
to link the Polyakov loop observables with those of the gauge-fixed
correlators~\cite{Fischer:2008uz,Maas:2011se}.

In the following, we investigate how different model assumptions of the
gauge-fixed correlators affect the fluctuations of the Polyakov loop.

\subsection{Susceptibilities and masses of Cartan angles}

We choose to focus on the physical case of $N_c=3$.
This case is unique in the sense that the two Cartan angles
$\gamma_{1,2}$ can be directly identified with the two degrees of freedom of
the Polyakov loop, i.e., $X, Y$. The (2 x 2) Jacobian matrix allows the
translation between $(\gamma_1, \gamma_2) \leftrightarrow (X, Y)$:

\begin{equation}
    \begin{split}
    J &= \frac{\partial \, \{X, Y\}}{\partial \, \{\gamma_1, \gamma_2 \}} \\
        J_{11} &= \frac{1}{3} \, \left( -\sin \frac{2 \gamma_1 + \gamma_2}{2} -
        \sin \frac{2 \gamma_1 - \gamma_2}{2} \right), \\
        J_{12} &= \frac{1}{3} \, \left( - \frac{1}{2} \, \sin \frac{2\gamma_1 +
        \gamma_2}{2} +  \frac{1}{2} \, \sin \frac{2\gamma_1 - \gamma_2}{2} -
        \sin \gamma_2 \right), \\
        J_{21} &= \frac{1}{3} \, \left( \cos \frac{2\gamma_1 + \gamma_2}{2} -
        \cos \frac{2\gamma_1 - \gamma_2}{2} \right), \\
        J_{22} &= \frac{1}{3} \, \left( - \frac{1}{2} \, \cos \frac{2\gamma_1 +
        \gamma_2}{2} +  \frac{1}{2} \, \cos \frac{2\gamma_1 - \gamma_2}{2} - \cos \gamma_2 \right).
    \end{split}
\end{equation}
We stress that
studying the order parameter and its fluctuations along two independent
directions is mandated by the existence of two independent Cartan generators, 
both relevant to describing the gauge group $SU(3)$. 
Many existing works have neglected the imaginary direction in the potential, 
and hence the appropriate field derivatives cannot be performed.

We define the (dimensionless) curvature mass tensor for the Cartan angles
as~\cite{Weiss:1981ev}
\begin{equation}
    \label{eq:carmass}
        \bar{m}^2_{ij} = \frac{\partial^2 \, U(\gamma_1, \gamma_2)}{\partial
        \gamma_i \, \partial \gamma_j} \, \frac{1}{T^4}.
\end{equation}
The tensor elements are to be evaluated with values of $\gamma_1$ which minimize the
potential, and $\gamma_2 \rightarrow 0$. 
For the class of potentials considered the off-diagonal terms vanish.
The relation to the curvature masses associated with the $(X, Y)$
fields~\cite{rob_dof} is thus

\begin{equation}
    \begin{split}
        \bar{m}^2_{11} &= J_{11}^2 \, \bar{m}^2_{XX} + J_{21}^2 \,
        \bar{m}^2_{YY}, \\
        \bar{m}^2_{22} &= J_{12}^2 \, \bar{m}^2_{XX} + J_{22}^2 \,
        \bar{m}^2_{YY},
    \end{split}
\end{equation}
where

\begin{equation}
    \begin{split}
        \bar{m}^2_{XX} &= \frac{\partial^2 \, U}{\partial
        X \, \partial X} \, \frac{1}{T^4}, \\
        \bar{m}^2_{YY} &= \frac{\partial^2 \, U}{\partial
        Y \, \partial Y} \, \frac{1}{T^4}.
    \end{split}
\end{equation}
A further simplification comes from the fact that the Jacobian matrix, 
evaluated along the real line (arbitrary $\gamma_1$, $\gamma_2=0$), is also diagonal:

\begin{equation}
    \begin{split}
        J_{11} &= -\frac{2}{3} \, \sin \gamma_1 = -\frac{1}{\sqrt{3}} \,
        \sqrt{(1- \ell )( 1+3 \ell )}, \\
        J_{22} &= -\frac{2}{3} \, ( \sin \frac{\gamma_1}{2} )^2 = -\frac{1}{2}
        \, (1- \ell), \\
        J_{12} &= 0, \\
        J_{21} &= 0,
    \end{split}
\end{equation}
where $\ell = \frac{1}{3} \, ( 1 + 2 \, \cos \gamma_1 )$.
Note that in the confined phase $\gamma_1 \rightarrow {2 \pi}/{3}$,

\begin{equation}
    \begin{split}
        J_{11} &\rightarrow -1/\sqrt{3}, \\
        J_{22} &\rightarrow -1/2,
    \end{split}
\end{equation}
and in the deconfined phase $\gamma_1 \rightarrow 0$ they vanish as

\begin{equation}
    \begin{split}
        J_{11} &\rightarrow -\frac{2}{\sqrt{3}} \,
        \sqrt{(1- \ell )}, \\
        J_{22} &\rightarrow -\frac{1}{2} \, (1- \ell),
    \end{split}
\end{equation}
with $\ell \rightarrow 1$. Note that $ \vert J_{22} \vert \ll \vert J_{11}
\vert $ in this limit.
We thus obtain the following relation between the curvature masses of Cartan
angles and the Polyakov loop susceptibilities:

\begin{equation}
    \label{eq:msq2sus}
    \begin{split}
        \bar{m}^2_{11} &= \frac{4}{9} \, (\sin \gamma_1)^2 \, \bar{m}^2_{XX}, \\
        \bar{m}^2_{22} &= \frac{4}{9} \, (\sin \frac{\gamma_1}{2})^4  \,
        \bar{m}^2_{YY},
    \end{split}
\end{equation}
with the Polyakov loop susceptibilities identified as 

\begin{equation}
    \label{eq:sus2msq}
    \begin{split}
        (T^3 \, \chi_L) &= \frac{1}{\bar{m}^2_{XX}} =  \frac{ (1- \ell )( 1+3 \ell )
        }{3 \, \bar{m}^2_{11}}, \\
        (T^3 \, \chi_T) &= \frac{1}{\bar{m}^2_{YY}} = \frac{(1- \ell)^2}{4 \, \bar{m}^2_{22}} .
    \end{split}
\end{equation}
This is a useful relation linking the Polyakov loop observables to those based
on the Cartan angles. The latter can eventually be linked to $A_0$ and the
transverse gluons.
Note that such a clean separation of contributions to $T^3 \, \chi_{L, T}$ from
$\bar{m}^2_{ii}$ is only true for $N_c=3$. Each susceptibility generally
receives contributions from all Cartan curvature masses $\bar{m}^2_{ii}$.

We derive an analytic expression for these Cartan curvature masses at ultrahigh temperatures. 
The effective potential is expected to approach

\begin{equation}
    U(\gamma_1, \gamma_2) \approx 2 \, U_1(m_A=0).
\end{equation}
Using the exact result of the integral

\begin{equation}
    \begin{split}
        U_a(Q_a) &= T \, \int \frac{d^3 k}{(2 \pi)^3} \, \ln \, ( I -
        e^{i \, Q_a} \, e^{-\beta k} )  \\
        &= -\frac{T^4}{\pi^2} \, {\rm PolyLog}(4, e^{i \, Q_a}),
    \end{split}
\end{equation}
and the polynomial expansion of the PolyLog function 
(valid in the restricted range of $Q_a \in [0, \pi]$)

\begin{equation}
        \frac{U_a(Q_a) + U_a(-Q_a)}{T^4} = -\frac{\pi^2}{45} + \frac{Q_a^2}{6} - \frac{Q_a^3}{6 \pi}
        + \frac{Q_a^4}{24 \pi^2},
\end{equation}
we obtain the potential (see Table~\ref{tab:su3q} and Eq.~\eqref{eq:su3gamma})

\begin{equation}
    \begin{split}
        \frac{U(\gamma_1, \gamma_2)}{T^4} &\approx -\frac{16 \pi^2}{90} + 2 \, \gamma_1^2  +
    \frac{3}{2} \, \gamma_2^2 \\
    & + \frac{3 \, (4 \gamma_1^2 + 3  \gamma_2^2)^2}{32 \pi^2} -  \frac{20
    \gamma_1^3 + 27 \gamma_1 \gamma_2^2}{6 \pi}. 
    \end{split}
\end{equation}
The curvature masses~\eqref{eq:carmass} can be readily deduced:

\begin{equation}
    \label{eq:htlim}
    \begin{split}
    \bar{m}^2_{11} &\rightarrow 4, \\
    \bar{m}^2_{22} &\rightarrow 3.
    \end{split}
\end{equation}
It follows from Eq.~\eqref{eq:sus2msq} that while both susceptibilities
approach zero at high temperatures, with $\chi_T \ll \chi_L$, 
one can extract a finite limit for the curvature masses.
Note that if we take 

\begin{equation}
    \gamma_{1,2} \rightarrow r_{1,2} \, \beta g A_4,
\end{equation}
these curvature masses are related to the dimensionful $m_{A_4}$ via

\begin{equation}
    \label{eq:matching}
    \begin{split}
        \bar{m}^2_{ii} &= \frac{m_{A_4}^2}{g^2 T^2 r_i^2}, \\
        {m}^2_{A_4} &= \frac{\partial^2 \, U}{\partial
        A_4 \, \partial A_4}
    \end{split}
\end{equation}
for $i = 1, 2$. The fact that $\bar{m}^2_{ii}$ has a finite limit forces

\begin{equation}
    m_{A_4} \propto g T, 
\end{equation}
as expected for a Debye screening mass. 
We stress that $m_{A_4}$ should be distinguished from the effective gluon mass
$m_A$ introduced. 
The latter captures the infrared enhancement originated from the nonperturbative vacuum 
and exists even at vanishing temperature.

The behaviors of these curvature masses at low temperatures are lesser known. 
In the confined phase, $Z(3)$ symmetry requires~\cite{Lo:2013etb}, in addition to $ \langle
\hat{\ell}_F \rangle = 0 $,

\begin{equation}
    \begin{split}
        \langle {\hat{\ell}_F}^2 \rangle &= 0 \\
        \implies \langle (X^2 - Y^2) \rangle &= 0.
    \end{split}
\end{equation}
This means the two susceptibilities are equal in this phase.
It follows from Eq.~\eqref{eq:sus2msq} that 

\begin{equation}
    \label{eq:cons}
    \frac{\bar{m}^2_{22}}{\bar{m}^2_{11}} = \frac{3}{4}
\end{equation}
in the confined phase. Note that the same ratio is observed in the high temperature limit~\eqref{eq:htlim}.

Other than the constraint~\eqref{eq:cons} on the ratio, 
there is no restriction from symmetry concerning their magnitudes.
In the language of an effective model, 
they reflect a competition between the confining (ghost) and the deconfining
(glue) parts of the potential. And unlike $\langle \hat{\ell}_F \rangle$, 
they are finite and temperature dependent even in the confined phase.

\begin{figure}[!ht]
\centering
\includegraphics[width=\linewidth]{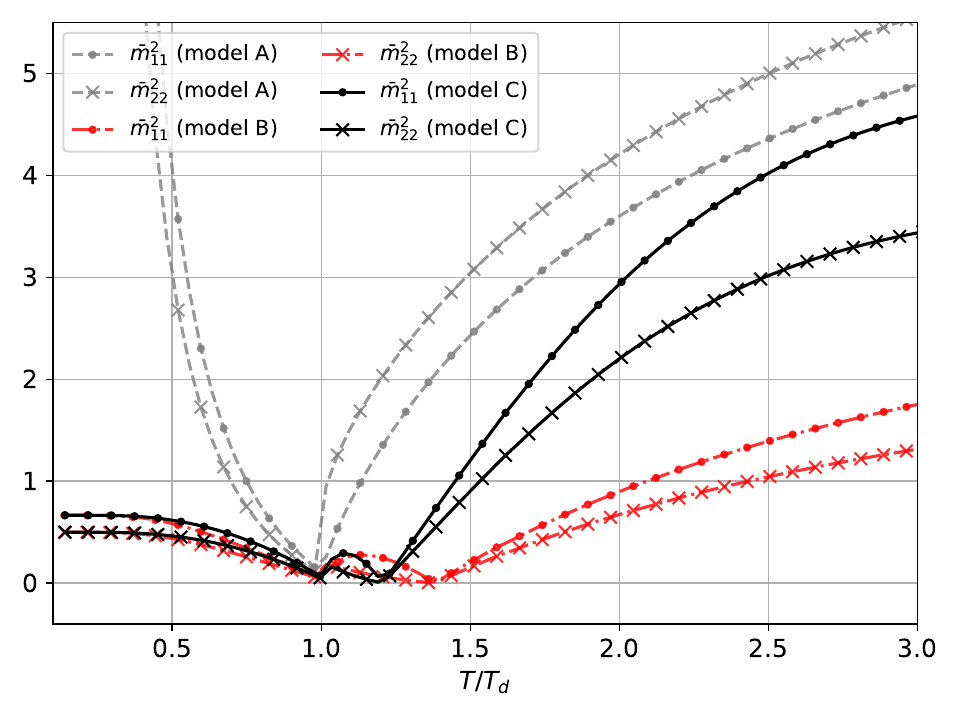}
        \caption{Curvature masses associated with the Cartan angles calculated
        for different models (see text).}
\label{fig5}
\end{figure}

To examine how the curvature masses associated with the
Cartan angles depend on the assumed properties of the gluons and ghosts, 
we compute the observables in the following arrangements of the effective potential:

\begin{itemize}

    \item model A: an invariant measure term~\eqref{eq:pot1} with two transverse gluons:

\begin{equation}
    \label{eq:modelA}
    U = -\frac{1}{2} \, b \, \ln H + 2 U_1(m_A)
\end{equation}

    \item model B: a ghost field term and three transverse gluons:

\begin{equation}
    \label{eq:modelB}
    U = -U_1(m_A=0) + 3 U_1(m_A)
\end{equation}

    \item model C: model B implemented with wave function renormalization
        effects discussed.

\end{itemize}
With no further tuning of model parameters, we obtain 
$T_d \approx (0.274, 0.274, 0.27)$ GeV for models A, B, and C.
The results of Cartan masses are shown in Fig.~\ref{fig5}.

We first report that Eq.~\eqref{eq:sus2msq} works: 
i.e., the same results of the susceptibilities are obtained when the curvature
tensor Eq.~\eqref{eq:xycurva} is directly constructed by taking the
appropriate $(X, Y)$-field derivatives on the potential derived in
Ref.~\cite{sasaki_pot}. 
This gives some confidence for the general applicability of
Eq.~\eqref{eq:dxy_op} for the general $N_c$ problem.

The most obvious feature of the curvature masses is the dip around $T_d$. 
Note that a very similar behavior is found for the $A_0$-gluon screening mass 
extracted from LQCD when studying the inverse of the longitudinal propagator~\cite{lqcd_1}. 
See also the discussion in Ref.~\cite{Maas:2011ez}.
In the effective model, this follows from the relation to
the Polyakov loop susceptibilities. 
While the gluon (and ghost) parameters employed are smooth, 
the discontinuity is inherited from minimizing the mean-field potential.
Note that a strong temperature dependence of these observables naturally arises 
without the need of introducing temperature dependent model parameters. 
In fact, in an improved scheme, the model parameters, including the additional
temperature dependence, should be determined self-consistently.

The high temperature limits~\eqref{eq:htlim} are approached very gradually: 
at $T \approx 30 \, T_d$ and from above (below) for models A, C (B). 
Note that model B has a known issue in the deconfined phase that the Polyakov loop reaches unity too rapidly and the model may not be reliable beyond
this point. 
Apparently, the existence of a secondary dip in the curvature masses in model B (also in
model C) also comes with this problem.
This does not happen to Model A, where the invariant measure term prevents
this problematic behavior.
It has been suggested that a two-loop calculation may remove this
artifact~\cite{2loop}.
It would be interesting to see the corresponding modification in the curvature
masses.

There is no strict theoretical constraint on the low temperature behaviors of these
curvature masses. 
The constraint~\eqref{eq:cons} on their ratio is verified in all cases.
What is clear from the effective model study is that they depend
strongly on the choice of the confining potential. 
This is particularly obvious in the $T \rightarrow 0$ limit:
In model A, they diverge as (see Eqs.~\eqref{eq:landau} and~\eqref{eq:msq2sus})

\begin{equation}
    \begin{split}
        \bar{m}^2_{11} &\rightarrow \frac{2 b}{T^3}, \\
        \bar{m}^2_{22} &\rightarrow \frac{3 b}{2 T^3}.
    \end{split}
\end{equation}
In model B we get instead the finite results:

\begin{equation}
    \begin{split}
        \bar{m}^2_{11} &\rightarrow \frac{2}{3}, \\
        \bar{m}^2_{22} &\rightarrow \frac{1}{2}.
    \end{split}
\end{equation}
The effect of wave function renormalization (model C), with the parameters chosen,
is found to be small at low temperatures, but becomes substantial close to and above $T_d$.

If we insist on imposing the matching condition~\eqref{eq:matching} 
and identify the $A_0$-gluon screening mass with $m_A$, 
we would obtain a $\propto \frac{1}{T^2}$ behavior for these curvature
masses.
It would thus be interesting to examine these observables with other gauge
choices~\cite{Langfeld:2004qs,Dudal:2007cw}: 
to see whether the differences are due to gauge artifacts,
and to gain insights in reliably describing the low temperature behavior of the Polyakov loop potential.

\begin{figure*}[!ht]
\centering
\includegraphics[width=0.49\linewidth]{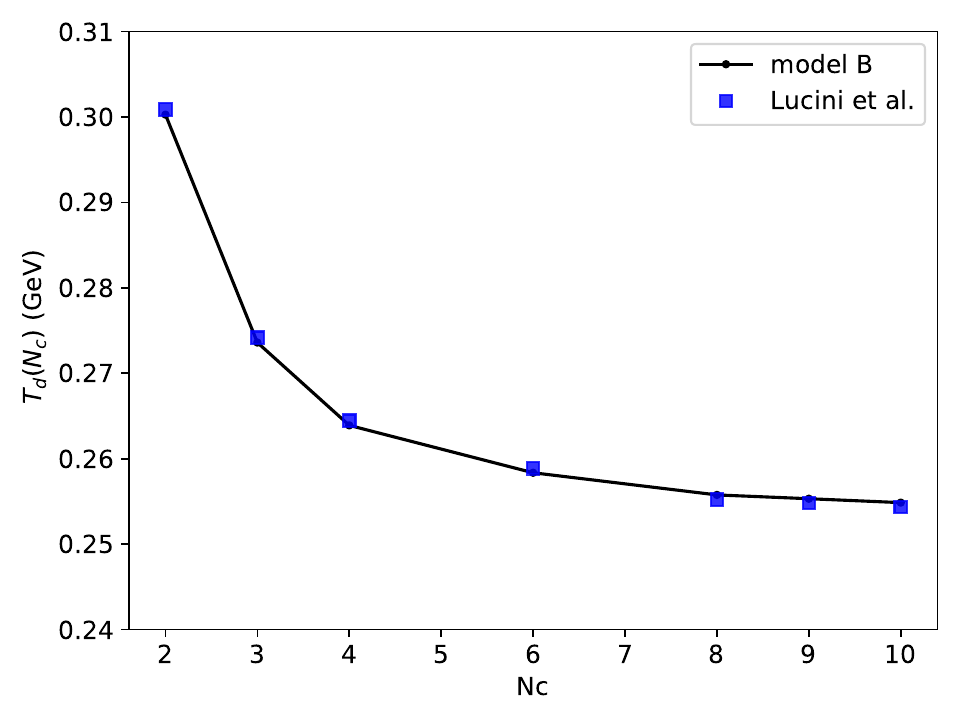}
\includegraphics[width=0.49\linewidth]{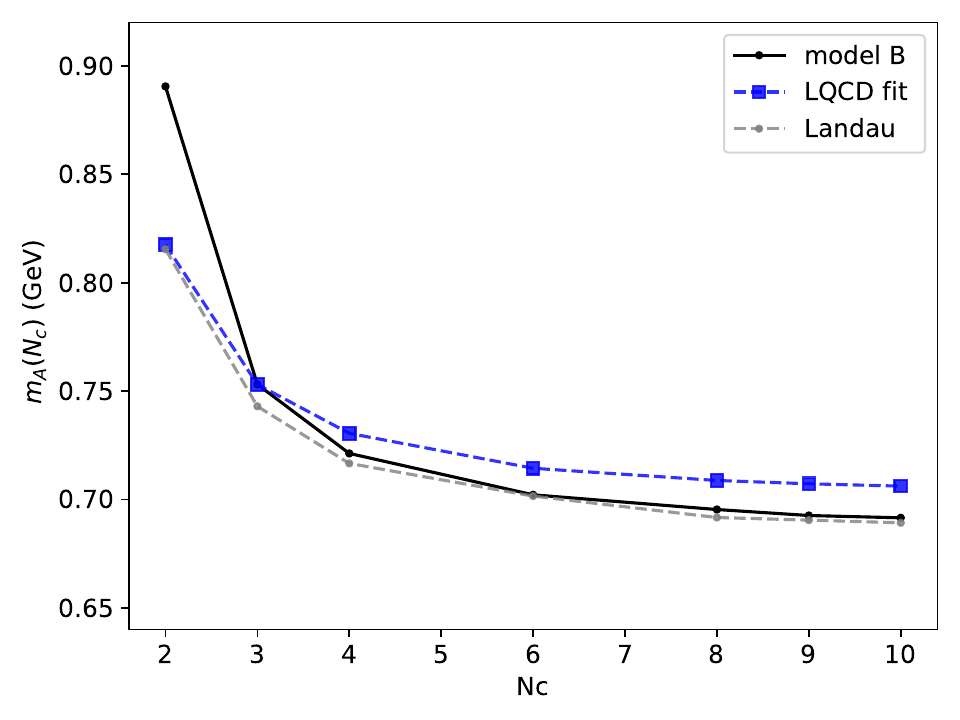}
        \caption{The critical temperatures $T_d$ of model B~\eqref{eq:modelB} (left)
        using the input masses $m_A(N_c)$ indicated in the right panel. 
        The latter are adjusted to fit the LQCD results on $T_d(N_c)$~\cite{lucini_td} and are compared to
        a fit to the LQCD of (half the) $0^{++}$ glueball
        mass~\cite{lucini_gb}. The gray dashed line shows 
        the result based on Landau parameter
        analysis~\eqref{eq:tdnc} ($\bar{u}_2$).}
\label{fig6}
\end{figure*}

\subsection{The appearance of glueballs}

Finally we speculate how the glueballs may enter the effective description.
In the current model a phenomenological gluon mass $m_A$ is introduced, 
which serves to suppress the gluon contribution (deconfining) in the potential at low temperatures.
This also sets the scale for the critical temperature $T_d$.

In Refs.~\cite{eric_gb,eric_cgauge}, 
a robust theoretical framework to introducing quasigluonic excitation is 
proposed via a constituent Fock space expansion. 
There, a nontrivial QCD vacuum, as in the Bardeen–Cooper–Schrieffer (BCS)
theory, is postulated, and with a Bogoliubov-Valatin
transformation the (massive) quasigluons are derived from the effective one-body
Hamiltonian. This mirrors the one-loop gluon potential considered here.

In addition, glueball spectra can be derived with the same Hamiltonian using
the two-gluon states built from these quasigluons. 
A key observation is that the lowest lying states receive most of their masses
from the quasigluons, i.e.,

\begin{equation}
    \label{eq:gbmass}
    m_{\rm GB} \approx 2 \, m_A, 
\end{equation}
e.g., $m_{\rm GB} \approx 1.7 (2.1)$ GeV for the lowest $0^{++} (0^{+-})$ state, 
with $m_A = 0.8$ GeV in Ref.~\cite{eric_gb}.
This naturally suggests a constituent model for the glueball states.
Neglecting their interactions with the Polyakov loops, we may consider free gas
of glueballs as an approximation for the $2 \rightarrow 2$ contribution to
the partition function.
~\footnote{This is similar to the case where a $\sigma$-meson is generated in
an Nambu–Jona-Lasinio model and
approximating its partial thermal pressure by a free Bose gas of mass
$m_\sigma \approx 2 M_Q$.}
See Ref.~\cite{Lacroix2013} for an 
elaborate treatment of thermal glueballs.

A nontrivial relation suggested by the effective model is a link between
$T_d(N_c)$ and $m_A(N_c)$. 
Model B is ideal for this illustration as 
$m_A$ is the only adjustable parameter of the model. 
Tuning $m_A$ to match the model $T_d$ with the LQCD results~\cite{lucini_td} for
various $N_c$'s, 
we extract the expected $N_c$ dependence of $m_A$. 
See Fig.~\ref{fig6}. 

The $m_A(N_c)$'s show a similar trend exhibited by a fit to LQCD $0^{++}$ glueball
masses~\cite{lucini_gb}. 
The fit employs the functional form

\begin{equation}
    m_{\rm GB}^{\rm LQCD}(N_c)/\sqrt{\sigma} \approx m_\infty + c/N_c^2
\end{equation}
with (dimensionless) parameters $m_\infty = 3.307, c = 2.187$, based on the
LQCD calculation in Ref.~\cite{lucini_gb} and 
fixing $c$ to the $N_c=3$ result.
We also take $\sigma = 0.18 \, {\rm GeV}^2$.

The general trend can easily be understood by studying the second Landau
parameter~\eqref{eq:landau}. For model B, it reads

\begin{equation}
    \label{eq:tdnc}
    \bar{u}_2 \approx \frac{N_c^2}{\pi^2} \, \left( 1 - 3 \, \frac{1}{2} \,
    (\frac{m_A}{T} )^2 \, K_2(m_A/T) \right).
\end{equation}
Solving for $m_A(N_c)$ from

\begin{equation}
    \label{eq:tdnc2}
\bar{u}_2(m_A(N_c), T=T_d(N_c)) = 0,
\end{equation}
we obtain the gray dashed line in Fig.~\ref{fig6} (right). 
Equation~\eqref{eq:tdnc} suggests that the leading $N_c$ dependence comes not from
the prefactor but from the $N_c$ dependence of
$T_d$.~\footnote{Relation~\eqref{eq:tdnc} assumes the Boltzmann approximation. This
may be justified for the massive gluons, but may not be the case for the ghost.
The corrections, however, are $N_c$ dependent. 
For $N_c=3$, this amounts to replacing $1 \rightarrow \frac{\pi^2}{9}
\approx 1.097$ in the right bracket. The corresponding result for $N_c=2$ is 
$1 \rightarrow \frac{\pi^2}{12} \approx 0.822$.}
While it is not surprising that the observables are related, 
the effective model offers a simple approximate connection such as~\eqref{eq:tdnc}.

\section{Conclusion}

We have examined the fluctuations of the order parameter, 
measured by the Polyakov loop susceptibilities in the $SU(N_c)$ pure gauge
theory,
using an effective potential built from one-loop expressions of the 
field determinants of gluon and ghost.
The connection between these observables with the Cartan angles and their
curvature masses are derived.
The latter can serve as a proxy for the $A_0$-gluon screening mass 
and are strongly influenced by the $Z(N_c)$ structure of the vacuum.

The Cartan curvature masses thus provide useful diagnostic information concerning
the competition of gluons and ghosts in the QCD medium. 
While we expect gauge invariance for all observables based on the Polyakov
loops, it is unlikely for the model potential in the current state to achieve
this goal. 
For example, we see that the predictions of these curvature masses 
depend strongly on the assumptions of the gluon and ghost propagators, 
and the choice of gauge.
Another essential limitation of the present model is that 
the propagators and wave function renormalizations we fitted 
are not LQCD computation in the background-field gauge.
A more constructive way to proceed is to explore the potential, 
and more generally the problem of how confinement manifests, in various gauges, 
and check whether there could be nontrivial relations among the model parameters 
such that the gauge dependence would be removed or reduced when computing physical observables.

A natural progression of this work is a consistent treatment of the finite
temperature behaviors of $m_A$ and the wave function renormalizations within
the model. A crude assessment gives the following competing effects: 
(1) slightly (Boltzmann-)suppressed deconfining force from transverse gluons due to 
an increased $m_A$ at finite temperatures, 
(2) an increase in confining force by the ghost due to the enhanced ghost form
factor,
(3) an increase in deconfining force from the longitudinal gluons due to the
characteristic dip in $m_{A_0}$.
It is possible that while $T_d$ would not be strongly affected, 
the curvature masses will be substantially modifed.

What is clear from the model study is that 
the longitudinal gluon propagator and the Cartan curvature masses are
connected (via Eqs. \eqref{eq:sus2msq} and \eqref{eq:matching}), and should be determined self-consistently in model calculations.
This makes for a more meaningful comparison with the finite temperature LQCD
data~\cite{Aouane:2011fv,lqcd_1,Bornyakov:2010nc,Bicudo:2017uyy}. 
For the transverse gluons, we find no evidence for a substantial change in their
masses across the phase transition, 
nor the need for the value to approach infinity in the confined
phase. In fact, they serve as constituents of the glueballs.

A possible future application of the relations between the Polyakov loop observables with those from the gluon
propagators could be in formulating a nonperturbative
renormalization scheme for the former as composite operators. 
This is a necessary first step to properly compare effective model results
with LQCD data of the Polyakov loops and the susceptibilities. 
This will be explored in a future work.

\section{Acknowledgments}
P.M.L thanks O. Oliveira for stimulating discussions.
We acknowledge the support by the Polish National Science Center (NCN) under Opus Grant No. 2018/31/B/ST2/01663. 
K.R. also acknowledges partial support of the Polish Ministry of Science and Higher Education.

\bibstyle{}
\bibliography{ref}

\begin{thebibliography}{65}%
\makeatletter
\providecommand \@ifxundefined [1]{%
 \@ifx{#1\undefined}
}%
\providecommand \@ifnum [1]{%
 \ifnum #1\expandafter \@firstoftwo
 \else \expandafter \@secondoftwo
 \fi
}%
\providecommand \@ifx [1]{%
 \ifx #1\expandafter \@firstoftwo
 \else \expandafter \@secondoftwo
 \fi
}%
\providecommand \natexlab [1]{#1}%
\providecommand \enquote  [1]{``#1''}%
\providecommand \bibnamefont  [1]{#1}%
\providecommand \bibfnamefont [1]{#1}%
\providecommand \citenamefont [1]{#1}%
\providecommand \href@noop [0]{\@secondoftwo}%
\providecommand \href [0]{\begingroup \@sanitize@url \@href}%
\providecommand \@href[1]{\@@startlink{#1}\@@href}%
\providecommand \@@href[1]{\endgroup#1\@@endlink}%
\providecommand \@sanitize@url [0]{\catcode `\\12\catcode `\$12\catcode
  `\&12\catcode `\#12\catcode `\^12\catcode `\_12\catcode `\%12\relax}%
\providecommand \@@startlink[1]{}%
\providecommand \@@endlink[0]{}%
\providecommand \url  [0]{\begingroup\@sanitize@url \@url }%
\providecommand \@url [1]{\endgroup\@href {#1}{\urlprefix }}%
\providecommand \urlprefix  [0]{URL }%
\providecommand \Eprint [0]{\href }%
\providecommand \doibase [0]{https://doi.org/}%
\providecommand \selectlanguage [0]{\@gobble}%
\providecommand \bibinfo  [0]{\@secondoftwo}%
\providecommand \bibfield  [0]{\@secondoftwo}%
\providecommand \translation [1]{[#1]}%
\providecommand \BibitemOpen [0]{}%
\providecommand \bibitemStop [0]{}%
\providecommand \bibitemNoStop [0]{.\EOS\space}%
\providecommand \EOS [0]{\spacefactor3000\relax}%
\providecommand \BibitemShut  [1]{\csname bibitem#1\endcsname}%
\let\auto@bib@innerbib\@empty
\bibitem [{\citenamefont {Boyd}\ \emph {et~al.}(1996)\citenamefont {Boyd},
  \citenamefont {Engels}, \citenamefont {Karsch}, \citenamefont {Laermann},
  \citenamefont {Legeland}, \citenamefont {Lutgemeier},\ and\ \citenamefont
  {Petersson}}]{Boyd:1996bx}%
  \BibitemOpen
  \bibfield  {author} {\bibinfo {author} {\bibfnamefont {G.}~\bibnamefont
  {Boyd}}, \bibinfo {author} {\bibfnamefont {J.}~\bibnamefont {Engels}},
  \bibinfo {author} {\bibfnamefont {F.}~\bibnamefont {Karsch}}, \bibinfo
  {author} {\bibfnamefont {E.}~\bibnamefont {Laermann}}, \bibinfo {author}
  {\bibfnamefont {C.}~\bibnamefont {Legeland}}, \bibinfo {author}
  {\bibfnamefont {M.}~\bibnamefont {Lutgemeier}},\ and\ \bibinfo {author}
  {\bibfnamefont {B.}~\bibnamefont {Petersson}},\ }\bibfield  {title} {\bibinfo
  {title} {{Thermodynamics of SU(3) lattice gauge theory}},\ }\href
  {https://doi.org/10.1016/0550-3213(96)00170-8} {\bibfield  {journal}
  {\bibinfo  {journal} {Nucl. Phys. B}\ }\textbf {\bibinfo {volume} {469}},\
  \bibinfo {pages} {419} (\bibinfo {year} {1996})},\ \Eprint
  {https://arxiv.org/abs/hep-lat/9602007} {arXiv:hep-lat/9602007} \BibitemShut
  {NoStop}%
\bibitem [{\citenamefont {Borsanyi}\ \emph {et~al.}(2012)\citenamefont
  {Borsanyi}, \citenamefont {Endrodi}, \citenamefont {Fodor}, \citenamefont
  {Katz},\ and\ \citenamefont {Szabo}}]{Borsanyi:2012ve}%
  \BibitemOpen
  \bibfield  {author} {\bibinfo {author} {\bibfnamefont {S.}~\bibnamefont
  {Borsanyi}}, \bibinfo {author} {\bibfnamefont {G.}~\bibnamefont {Endrodi}},
  \bibinfo {author} {\bibfnamefont {Z.}~\bibnamefont {Fodor}}, \bibinfo
  {author} {\bibfnamefont {S.~D.}\ \bibnamefont {Katz}},\ and\ \bibinfo
  {author} {\bibfnamefont {K.~K.}\ \bibnamefont {Szabo}},\ }\bibfield  {title}
  {\bibinfo {title} {{Precision SU(3) lattice thermodynamics for a large
  temperature range}},\ }\href {https://doi.org/10.1007/JHEP07(2012)056}
  {\bibfield  {journal} {\bibinfo  {journal} {JHEP}\ }\textbf {\bibinfo
  {volume} {07}},\ \bibinfo {pages} {056}},\ \Eprint
  {https://arxiv.org/abs/1204.6184} {arXiv:1204.6184 [hep-lat]} \BibitemShut
  {NoStop}%
\bibitem [{\citenamefont {Kaczmarek}\ \emph {et~al.}(2002)\citenamefont
  {Kaczmarek}, \citenamefont {Karsch}, \citenamefont {Petreczky},\ and\
  \citenamefont {Zantow}}]{Kaczmarek:2002mc}%
  \BibitemOpen
  \bibfield  {author} {\bibinfo {author} {\bibfnamefont {O.}~\bibnamefont
  {Kaczmarek}}, \bibinfo {author} {\bibfnamefont {F.}~\bibnamefont {Karsch}},
  \bibinfo {author} {\bibfnamefont {P.}~\bibnamefont {Petreczky}},\ and\
  \bibinfo {author} {\bibfnamefont {F.}~\bibnamefont {Zantow}},\ }\bibfield
  {title} {\bibinfo {title} {{Heavy quark anti-quark free energy and the
  renormalized Polyakov loop}},\ }\href
  {https://doi.org/10.1016/S0370-2693(02)02415-2} {\bibfield  {journal}
  {\bibinfo  {journal} {Phys. Lett. B}\ }\textbf {\bibinfo {volume} {543}},\
  \bibinfo {pages} {41} (\bibinfo {year} {2002})},\ \Eprint
  {https://arxiv.org/abs/hep-lat/0207002} {arXiv:hep-lat/0207002} \BibitemShut
  {NoStop}%
\bibitem [{\citenamefont {Fukushima}\ and\ \citenamefont
  {Sasaki}(2013)}]{Fukushima:2013rx}%
  \BibitemOpen
  \bibfield  {author} {\bibinfo {author} {\bibfnamefont {K.}~\bibnamefont
  {Fukushima}}\ and\ \bibinfo {author} {\bibfnamefont {C.}~\bibnamefont
  {Sasaki}},\ }\bibfield  {title} {\bibinfo {title} {{The phase diagram of
  nuclear and quark matter at high baryon density}},\ }\href
  {https://doi.org/10.1016/j.ppnp.2013.05.003} {\bibfield  {journal} {\bibinfo
  {journal} {Prog. Part. Nucl. Phys.}\ }\textbf {\bibinfo {volume} {72}},\
  \bibinfo {pages} {99} (\bibinfo {year} {2013})},\ \Eprint
  {https://arxiv.org/abs/1301.6377} {arXiv:1301.6377 [hep-ph]} \BibitemShut
  {NoStop}%
\bibitem [{\citenamefont {Fukushima}\ and\ \citenamefont
  {Skokov}(2017)}]{Fukushima:2017csk}%
  \BibitemOpen
  \bibfield  {author} {\bibinfo {author} {\bibfnamefont {K.}~\bibnamefont
  {Fukushima}}\ and\ \bibinfo {author} {\bibfnamefont {V.}~\bibnamefont
  {Skokov}},\ }\bibfield  {title} {\bibinfo {title} {{Polyakov loop modeling
  for hot QCD}},\ }\href {https://doi.org/10.1016/j.ppnp.2017.05.002}
  {\bibfield  {journal} {\bibinfo  {journal} {Prog. Part. Nucl. Phys.}\
  }\textbf {\bibinfo {volume} {96}},\ \bibinfo {pages} {154} (\bibinfo {year}
  {2017})},\ \Eprint {https://arxiv.org/abs/1705.00718} {arXiv:1705.00718
  [hep-ph]} \BibitemShut {NoStop}%
\bibitem [{\citenamefont {Andersen}\ \emph {et~al.}(2016)\citenamefont
  {Andersen}, \citenamefont {Naylor},\ and\ \citenamefont
  {Tranberg}}]{Andersen:2014xxa}%
  \BibitemOpen
  \bibfield  {author} {\bibinfo {author} {\bibfnamefont {J.~O.}\ \bibnamefont
  {Andersen}}, \bibinfo {author} {\bibfnamefont {W.~R.}\ \bibnamefont
  {Naylor}},\ and\ \bibinfo {author} {\bibfnamefont {A.}~\bibnamefont
  {Tranberg}},\ }\bibfield  {title} {\bibinfo {title} {{Phase diagram of QCD in
  a magnetic field: A review}},\ }\href
  {https://doi.org/10.1103/RevModPhys.88.025001} {\bibfield  {journal}
  {\bibinfo  {journal} {Rev. Mod. Phys.}\ }\textbf {\bibinfo {volume} {88}},\
  \bibinfo {pages} {025001} (\bibinfo {year} {2016})},\ \Eprint
  {https://arxiv.org/abs/1411.7176} {arXiv:1411.7176 [hep-ph]} \BibitemShut
  {NoStop}%
\bibitem [{\citenamefont {Bruckmann}\ \emph {et~al.}(2013)\citenamefont
  {Bruckmann}, \citenamefont {Endrodi},\ and\ \citenamefont
  {Kovacs}}]{Bruckmann:2013oba}%
  \BibitemOpen
  \bibfield  {author} {\bibinfo {author} {\bibfnamefont {F.}~\bibnamefont
  {Bruckmann}}, \bibinfo {author} {\bibfnamefont {G.}~\bibnamefont {Endrodi}},\
  and\ \bibinfo {author} {\bibfnamefont {T.~G.}\ \bibnamefont {Kovacs}},\
  }\bibfield  {title} {\bibinfo {title} {{Inverse magnetic catalysis and the
  Polyakov loop}},\ }\href {https://doi.org/10.1007/JHEP04(2013)112} {\bibfield
   {journal} {\bibinfo  {journal} {JHEP}\ }\textbf {\bibinfo {volume} {04}},\
  \bibinfo {pages} {112}},\ \Eprint {https://arxiv.org/abs/1303.3972}
  {arXiv:1303.3972 [hep-lat]} \BibitemShut {NoStop}%
\bibitem [{\citenamefont {Fraga}\ \emph {et~al.}(2014)\citenamefont {Fraga},
  \citenamefont {Mintz},\ and\ \citenamefont
  {Schaffner-Bielich}}]{Fraga:2013ova}%
  \BibitemOpen
  \bibfield  {author} {\bibinfo {author} {\bibfnamefont {E.~S.}\ \bibnamefont
  {Fraga}}, \bibinfo {author} {\bibfnamefont {B.~W.}\ \bibnamefont {Mintz}},\
  and\ \bibinfo {author} {\bibfnamefont {J.}~\bibnamefont
  {Schaffner-Bielich}},\ }\bibfield  {title} {\bibinfo {title} {{A search for
  inverse magnetic catalysis in thermal quark-meson models}},\ }\href
  {https://doi.org/10.1016/j.physletb.2014.02.028} {\bibfield  {journal}
  {\bibinfo  {journal} {Phys. Lett. B}\ }\textbf {\bibinfo {volume} {731}},\
  \bibinfo {pages} {154} (\bibinfo {year} {2014})},\ \Eprint
  {https://arxiv.org/abs/1311.3964} {arXiv:1311.3964 [hep-ph]} \BibitemShut
  {NoStop}%
\bibitem [{\citenamefont {Pagura}\ \emph {et~al.}(2017)\citenamefont {Pagura},
  \citenamefont {Gomez~Dumm}, \citenamefont {Noguera},\ and\ \citenamefont
  {Scoccola}}]{Pagura:2016pwr}%
  \BibitemOpen
  \bibfield  {author} {\bibinfo {author} {\bibfnamefont {V.~P.}\ \bibnamefont
  {Pagura}}, \bibinfo {author} {\bibfnamefont {D.}~\bibnamefont {Gomez~Dumm}},
  \bibinfo {author} {\bibfnamefont {S.}~\bibnamefont {Noguera}},\ and\ \bibinfo
  {author} {\bibfnamefont {N.~N.}\ \bibnamefont {Scoccola}},\ }\bibfield
  {title} {\bibinfo {title} {{Magnetic catalysis and inverse magnetic catalysis
  in nonlocal chiral quark models}},\ }\href
  {https://doi.org/10.1103/PhysRevD.95.034013} {\bibfield  {journal} {\bibinfo
  {journal} {Phys. Rev. D}\ }\textbf {\bibinfo {volume} {95}},\ \bibinfo
  {pages} {034013} (\bibinfo {year} {2017})},\ \Eprint
  {https://arxiv.org/abs/1609.02025} {arXiv:1609.02025 [hep-ph]} \BibitemShut
  {NoStop}%
\bibitem [{\citenamefont {Lo}\ \emph {et~al.}(2018)\citenamefont {Lo},
  \citenamefont {Szyma\'nski}, \citenamefont {Redlich},\ and\ \citenamefont
  {Sasaki}}]{Lo:2018wdo}%
  \BibitemOpen
  \bibfield  {author} {\bibinfo {author} {\bibfnamefont {P.~M.}\ \bibnamefont
  {Lo}}, \bibinfo {author} {\bibfnamefont {M.}~\bibnamefont {Szyma\'nski}},
  \bibinfo {author} {\bibfnamefont {K.}~\bibnamefont {Redlich}},\ and\ \bibinfo
  {author} {\bibfnamefont {C.}~\bibnamefont {Sasaki}},\ }\bibfield  {title}
  {\bibinfo {title} {{Polyakov loop fluctuations in the presence of external
  fields}},\ }\href {https://doi.org/10.1103/PhysRevD.97.114006} {\bibfield
  {journal} {\bibinfo  {journal} {Phys. Rev. D}\ }\textbf {\bibinfo {volume}
  {97}},\ \bibinfo {pages} {114006} (\bibinfo {year} {2018})},\ \Eprint
  {https://arxiv.org/abs/1801.08040} {arXiv:1801.08040 [hep-ph]} \BibitemShut
  {NoStop}%
\bibitem [{\citenamefont {Lo}\ \emph {et~al.}(2020)\citenamefont {Lo},
  \citenamefont {Szyma\'nski}, \citenamefont {Sasaki},\ and\ \citenamefont
  {Redlich}}]{Lo:2020ptj}%
  \BibitemOpen
  \bibfield  {author} {\bibinfo {author} {\bibfnamefont {P.~M.}\ \bibnamefont
  {Lo}}, \bibinfo {author} {\bibfnamefont {M.}~\bibnamefont {Szyma\'nski}},
  \bibinfo {author} {\bibfnamefont {C.}~\bibnamefont {Sasaki}},\ and\ \bibinfo
  {author} {\bibfnamefont {K.}~\bibnamefont {Redlich}},\ }\bibfield  {title}
  {\bibinfo {title} {{Deconfinement in the presence of a strong magnetic
  field}},\ }\href {https://doi.org/10.1103/PhysRevD.102.034024} {\bibfield
  {journal} {\bibinfo  {journal} {Phys. Rev. D}\ }\textbf {\bibinfo {volume}
  {102}},\ \bibinfo {pages} {034024} (\bibinfo {year} {2020})},\ \Eprint
  {https://arxiv.org/abs/2004.04138} {arXiv:2004.04138 [hep-ph]} \BibitemShut
  {NoStop}%
\bibitem [{\citenamefont {Ratti}\ \emph {et~al.}(2006)\citenamefont {Ratti},
  \citenamefont {Thaler},\ and\ \citenamefont {Weise}}]{Ratti:2005jh}%
  \BibitemOpen
  \bibfield  {author} {\bibinfo {author} {\bibfnamefont {C.}~\bibnamefont
  {Ratti}}, \bibinfo {author} {\bibfnamefont {M.~A.}\ \bibnamefont {Thaler}},\
  and\ \bibinfo {author} {\bibfnamefont {W.}~\bibnamefont {Weise}},\ }\bibfield
   {title} {\bibinfo {title} {{Phases of QCD: Lattice thermodynamics and a
  field theoretical model}},\ }\href
  {https://doi.org/10.1103/PhysRevD.73.014019} {\bibfield  {journal} {\bibinfo
  {journal} {Phys. Rev. D}\ }\textbf {\bibinfo {volume} {73}},\ \bibinfo
  {pages} {014019} (\bibinfo {year} {2006})},\ \Eprint
  {https://arxiv.org/abs/hep-ph/0506234} {arXiv:hep-ph/0506234} \BibitemShut
  {NoStop}%
\bibitem [{\citenamefont {Dumitru}\ \emph {et~al.}(2012)\citenamefont
  {Dumitru}, \citenamefont {Guo}, \citenamefont {Hidaka}, \citenamefont
  {Altes},\ and\ \citenamefont {Pisarski}}]{mat_model_1}%
  \BibitemOpen
  \bibfield  {author} {\bibinfo {author} {\bibfnamefont {A.}~\bibnamefont
  {Dumitru}}, \bibinfo {author} {\bibfnamefont {Y.}~\bibnamefont {Guo}},
  \bibinfo {author} {\bibfnamefont {Y.}~\bibnamefont {Hidaka}}, \bibinfo
  {author} {\bibfnamefont {C.~P.}\ \bibnamefont {Altes}},\ and\ \bibinfo
  {author} {\bibfnamefont {R.~D.}\ \bibnamefont {Pisarski}},\ }\bibfield
  {title} {\bibinfo {title} {{Effective Matrix Model for Deconfinement in Pure
  Gauge Theories}},\ }\href {https://doi.org/10.1103/PhysRevD.86.105017}
  {\bibfield  {journal} {\bibinfo  {journal} {Phys. Rev. D}\ }\textbf {\bibinfo
  {volume} {86}},\ \bibinfo {pages} {105017} (\bibinfo {year} {2012})},\
  \Eprint {https://arxiv.org/abs/1205.0137} {arXiv:1205.0137 [hep-ph]}
  \BibitemShut {NoStop}%
\bibitem [{\citenamefont {Lo}\ \emph {et~al.}(2013{\natexlab{a}})\citenamefont
  {Lo}, \citenamefont {Friman}, \citenamefont {Kaczmarek}, \citenamefont
  {Redlich},\ and\ \citenamefont {Sasaki}}]{Lo:2013hla}%
  \BibitemOpen
  \bibfield  {author} {\bibinfo {author} {\bibfnamefont {P.~M.}\ \bibnamefont
  {Lo}}, \bibinfo {author} {\bibfnamefont {B.}~\bibnamefont {Friman}}, \bibinfo
  {author} {\bibfnamefont {O.}~\bibnamefont {Kaczmarek}}, \bibinfo {author}
  {\bibfnamefont {K.}~\bibnamefont {Redlich}},\ and\ \bibinfo {author}
  {\bibfnamefont {C.}~\bibnamefont {Sasaki}},\ }\bibfield  {title} {\bibinfo
  {title} {{Polyakov loop fluctuations in SU(3) lattice gauge theory and an
  effective gluon potential}},\ }\href
  {https://doi.org/10.1103/PhysRevD.88.074502} {\bibfield  {journal} {\bibinfo
  {journal} {Phys. Rev. D}\ }\textbf {\bibinfo {volume} {88}},\ \bibinfo
  {pages} {074502} (\bibinfo {year} {2013}{\natexlab{a}})},\ \Eprint
  {https://arxiv.org/abs/1307.5958} {arXiv:1307.5958 [hep-lat]} \BibitemShut
  {NoStop}%
\bibitem [{\citenamefont {Reinosa}\ \emph {et~al.}(2015)\citenamefont
  {Reinosa}, \citenamefont {Serreau}, \citenamefont {Tissier},\ and\
  \citenamefont {Wschebor}}]{Reinosa:2014ooa}%
  \BibitemOpen
  \bibfield  {author} {\bibinfo {author} {\bibfnamefont {U.}~\bibnamefont
  {Reinosa}}, \bibinfo {author} {\bibfnamefont {J.}~\bibnamefont {Serreau}},
  \bibinfo {author} {\bibfnamefont {M.}~\bibnamefont {Tissier}},\ and\ \bibinfo
  {author} {\bibfnamefont {N.}~\bibnamefont {Wschebor}},\ }\bibfield  {title}
  {\bibinfo {title} {{Deconfinement transition in SU($N$) theories from
  perturbation theory}},\ }\href
  {https://doi.org/10.1016/j.physletb.2015.01.006} {\bibfield  {journal}
  {\bibinfo  {journal} {Phys. Lett. B}\ }\textbf {\bibinfo {volume} {742}},\
  \bibinfo {pages} {61} (\bibinfo {year} {2015})},\ \Eprint
  {https://arxiv.org/abs/1407.6469} {arXiv:1407.6469 [hep-ph]} \BibitemShut
  {NoStop}%
\bibitem [{\citenamefont {Braun}\ \emph
  {et~al.}(2010{\natexlab{a}})\citenamefont {Braun}, \citenamefont {Gies},\
  and\ \citenamefont {Pawlowski}}]{Braun:2007bx}%
  \BibitemOpen
  \bibfield  {author} {\bibinfo {author} {\bibfnamefont {J.}~\bibnamefont
  {Braun}}, \bibinfo {author} {\bibfnamefont {H.}~\bibnamefont {Gies}},\ and\
  \bibinfo {author} {\bibfnamefont {J.~M.}\ \bibnamefont {Pawlowski}},\
  }\bibfield  {title} {\bibinfo {title} {{Quark Confinement from Color
  Confinement}},\ }\href {https://doi.org/10.1016/j.physletb.2010.01.009}
  {\bibfield  {journal} {\bibinfo  {journal} {Phys. Lett. B}\ }\textbf
  {\bibinfo {volume} {684}},\ \bibinfo {pages} {262} (\bibinfo {year}
  {2010}{\natexlab{a}})},\ \Eprint {https://arxiv.org/abs/0708.2413}
  {arXiv:0708.2413 [hep-th]} \BibitemShut {NoStop}%
\bibitem [{\citenamefont {Meisinger}\ \emph {et~al.}(2002)\citenamefont
  {Meisinger}, \citenamefont {Miller},\ and\ \citenamefont
  {Ogilvie}}]{Meisinger:2001cq}%
  \BibitemOpen
  \bibfield  {author} {\bibinfo {author} {\bibfnamefont {P.~N.}\ \bibnamefont
  {Meisinger}}, \bibinfo {author} {\bibfnamefont {T.~R.}\ \bibnamefont
  {Miller}},\ and\ \bibinfo {author} {\bibfnamefont {M.~C.}\ \bibnamefont
  {Ogilvie}},\ }\bibfield  {title} {\bibinfo {title} {{Phenomenological
  equations of state for the quark gluon plasma}},\ }\href
  {https://doi.org/10.1103/PhysRevD.65.034009} {\bibfield  {journal} {\bibinfo
  {journal} {Phys. Rev. D}\ }\textbf {\bibinfo {volume} {65}},\ \bibinfo
  {pages} {034009} (\bibinfo {year} {2002})},\ \Eprint
  {https://arxiv.org/abs/hep-ph/0108009} {arXiv:hep-ph/0108009} \BibitemShut
  {NoStop}%
\bibitem [{\citenamefont {Meisinger}\ \emph {et~al.}(2004)\citenamefont
  {Meisinger}, \citenamefont {Ogilvie},\ and\ \citenamefont
  {Miller}}]{Meisinger:2003id}%
  \BibitemOpen
  \bibfield  {author} {\bibinfo {author} {\bibfnamefont {P.~N.}\ \bibnamefont
  {Meisinger}}, \bibinfo {author} {\bibfnamefont {M.~C.}\ \bibnamefont
  {Ogilvie}},\ and\ \bibinfo {author} {\bibfnamefont {T.~R.}\ \bibnamefont
  {Miller}},\ }\bibfield  {title} {\bibinfo {title} {{Gluon quasiparticles and
  the polyakov loop}},\ }\href {https://doi.org/10.1016/j.physletb.2004.02.009}
  {\bibfield  {journal} {\bibinfo  {journal} {Phys. Lett. B}\ }\textbf
  {\bibinfo {volume} {585}},\ \bibinfo {pages} {149} (\bibinfo {year}
  {2004})},\ \Eprint {https://arxiv.org/abs/hep-ph/0312272}
  {arXiv:hep-ph/0312272} \BibitemShut {NoStop}%
\bibitem [{\citenamefont {Alba}\ \emph {et~al.}(2014)\citenamefont {Alba},
  \citenamefont {Alberico}, \citenamefont {Bluhm}, \citenamefont {Greco},
  \citenamefont {Ratti},\ and\ \citenamefont {Ruggieri}}]{gen_1}%
  \BibitemOpen
  \bibfield  {author} {\bibinfo {author} {\bibfnamefont {P.}~\bibnamefont
  {Alba}}, \bibinfo {author} {\bibfnamefont {W.}~\bibnamefont {Alberico}},
  \bibinfo {author} {\bibfnamefont {M.}~\bibnamefont {Bluhm}}, \bibinfo
  {author} {\bibfnamefont {V.}~\bibnamefont {Greco}}, \bibinfo {author}
  {\bibfnamefont {C.}~\bibnamefont {Ratti}},\ and\ \bibinfo {author}
  {\bibfnamefont {M.}~\bibnamefont {Ruggieri}},\ }\bibfield  {title} {\bibinfo
  {title} {{Polyakov loop and gluon quasiparticles: A self-consistent approach
  to Yang\textendash{}Mills thermodynamics}},\ }\href
  {https://doi.org/10.1016/j.nuclphysa.2014.11.011} {\bibfield  {journal}
  {\bibinfo  {journal} {Nucl. Phys. A}\ }\textbf {\bibinfo {volume} {934}},\
  \bibinfo {pages} {41} (\bibinfo {year} {2014})},\ \Eprint
  {https://arxiv.org/abs/1402.6213} {arXiv:1402.6213 [hep-ph]} \BibitemShut
  {NoStop}%
\bibitem [{\citenamefont {Bannur}(2007)}]{Bannur:2006js}%
  \BibitemOpen
  \bibfield  {author} {\bibinfo {author} {\bibfnamefont {V.~M.}\ \bibnamefont
  {Bannur}},\ }\bibfield  {title} {\bibinfo {title} {{Self-consistent
  quasiparticle model for quark-gluon plasma}},\ }\href
  {https://doi.org/10.1103/PhysRevC.75.044905} {\bibfield  {journal} {\bibinfo
  {journal} {Phys. Rev. C}\ }\textbf {\bibinfo {volume} {75}},\ \bibinfo
  {pages} {044905} (\bibinfo {year} {2007})},\ \Eprint
  {https://arxiv.org/abs/hep-ph/0609188} {arXiv:hep-ph/0609188} \BibitemShut
  {NoStop}%
\bibitem [{\citenamefont {Braun}\ \emph
  {et~al.}(2010{\natexlab{b}})\citenamefont {Braun}, \citenamefont {Eichhorn},
  \citenamefont {Gies},\ and\ \citenamefont {Pawlowski}}]{referee4}%
  \BibitemOpen
  \bibfield  {author} {\bibinfo {author} {\bibfnamefont {J.}~\bibnamefont
  {Braun}}, \bibinfo {author} {\bibfnamefont {A.}~\bibnamefont {Eichhorn}},
  \bibinfo {author} {\bibfnamefont {H.}~\bibnamefont {Gies}},\ and\ \bibinfo
  {author} {\bibfnamefont {J.~M.}\ \bibnamefont {Pawlowski}},\ }\bibfield
  {title} {\bibinfo {title} {{On the Nature of the Phase Transition in SU(N),
  Sp(2) and E(7) Yang-Mills theory}},\ }\href
  {https://doi.org/10.1140/epjc/s10052-010-1485-1} {\bibfield  {journal}
  {\bibinfo  {journal} {Eur. Phys. J. C}\ }\textbf {\bibinfo {volume} {70}},\
  \bibinfo {pages} {689} (\bibinfo {year} {2010}{\natexlab{b}})},\ \Eprint
  {https://arxiv.org/abs/1007.2619} {arXiv:1007.2619 [hep-ph]} \BibitemShut
  {NoStop}%
\bibitem [{\citenamefont {Sasaki}\ and\ \citenamefont
  {Redlich}(2012)}]{sasaki_pot}%
  \BibitemOpen
  \bibfield  {author} {\bibinfo {author} {\bibfnamefont {C.}~\bibnamefont
  {Sasaki}}\ and\ \bibinfo {author} {\bibfnamefont {K.}~\bibnamefont
  {Redlich}},\ }\bibfield  {title} {\bibinfo {title} {{An Effective gluon
  potential and hybrid approach to Yang-Mills thermodynamics}},\ }\href
  {https://doi.org/10.1103/PhysRevD.86.014007} {\bibfield  {journal} {\bibinfo
  {journal} {Phys. Rev. D}\ }\textbf {\bibinfo {volume} {86}},\ \bibinfo
  {pages} {014007} (\bibinfo {year} {2012})},\ \Eprint
  {https://arxiv.org/abs/1204.4330} {arXiv:1204.4330 [hep-ph]} \BibitemShut
  {NoStop}%
\bibitem [{\citenamefont {Fukushima}\ and\ \citenamefont
  {Kashiwa}(2013)}]{Fukushima:2012qa}%
  \BibitemOpen
  \bibfield  {author} {\bibinfo {author} {\bibfnamefont {K.}~\bibnamefont
  {Fukushima}}\ and\ \bibinfo {author} {\bibfnamefont {K.}~\bibnamefont
  {Kashiwa}},\ }\bibfield  {title} {\bibinfo {title} {{Polyakov loop and QCD
  thermodynamics from the gluon and ghost propagators}},\ }\href
  {https://doi.org/10.1016/j.physletb.2013.05.037} {\bibfield  {journal}
  {\bibinfo  {journal} {Phys. Lett. B}\ }\textbf {\bibinfo {volume} {723}},\
  \bibinfo {pages} {360} (\bibinfo {year} {2013})},\ \Eprint
  {https://arxiv.org/abs/1206.0685} {arXiv:1206.0685 [hep-ph]} \BibitemShut
  {NoStop}%
\bibitem [{\citenamefont {Lo}\ \emph {et~al.}(2013{\natexlab{b}})\citenamefont
  {Lo}, \citenamefont {Friman}, \citenamefont {Kaczmarek}, \citenamefont
  {Redlich},\ and\ \citenamefont {Sasaki}}]{Lo:2013etb}%
  \BibitemOpen
  \bibfield  {author} {\bibinfo {author} {\bibfnamefont {P.~M.}\ \bibnamefont
  {Lo}}, \bibinfo {author} {\bibfnamefont {B.}~\bibnamefont {Friman}}, \bibinfo
  {author} {\bibfnamefont {O.}~\bibnamefont {Kaczmarek}}, \bibinfo {author}
  {\bibfnamefont {K.}~\bibnamefont {Redlich}},\ and\ \bibinfo {author}
  {\bibfnamefont {C.}~\bibnamefont {Sasaki}},\ }\bibfield  {title} {\bibinfo
  {title} {{Probing Deconfinement with Polyakov Loop Susceptibilities}},\
  }\href {https://doi.org/10.1103/PhysRevD.88.014506} {\bibfield  {journal}
  {\bibinfo  {journal} {Phys. Rev. D}\ }\textbf {\bibinfo {volume} {88}},\
  \bibinfo {pages} {014506} (\bibinfo {year} {2013}{\natexlab{b}})},\ \Eprint
  {https://arxiv.org/abs/1306.5094} {arXiv:1306.5094 [hep-lat]} \BibitemShut
  {NoStop}%
\bibitem [{\citenamefont {Sasaki}\ \emph {et~al.}(2007)\citenamefont {Sasaki},
  \citenamefont {Friman},\ and\ \citenamefont {Redlich}}]{Sasaki:2006ww}%
  \BibitemOpen
  \bibfield  {author} {\bibinfo {author} {\bibfnamefont {C.}~\bibnamefont
  {Sasaki}}, \bibinfo {author} {\bibfnamefont {B.}~\bibnamefont {Friman}},\
  and\ \bibinfo {author} {\bibfnamefont {K.}~\bibnamefont {Redlich}},\
  }\bibfield  {title} {\bibinfo {title} {{Susceptibilities and the Phase
  Structure of a Chiral Model with Polyakov Loops}},\ }\href
  {https://doi.org/10.1103/PhysRevD.75.074013} {\bibfield  {journal} {\bibinfo
  {journal} {Phys. Rev. D}\ }\textbf {\bibinfo {volume} {75}},\ \bibinfo
  {pages} {074013} (\bibinfo {year} {2007})},\ \Eprint
  {https://arxiv.org/abs/hep-ph/0611147} {arXiv:hep-ph/0611147} \BibitemShut
  {NoStop}%
\bibitem [{\citenamefont {Lo}\ \emph {et~al.}(2014)\citenamefont {Lo},
  \citenamefont {Friman},\ and\ \citenamefont {Redlich}}]{Lo:2014vba}%
  \BibitemOpen
  \bibfield  {author} {\bibinfo {author} {\bibfnamefont {P.~M.}\ \bibnamefont
  {Lo}}, \bibinfo {author} {\bibfnamefont {B.}~\bibnamefont {Friman}},\ and\
  \bibinfo {author} {\bibfnamefont {K.}~\bibnamefont {Redlich}},\ }\bibfield
  {title} {\bibinfo {title} {{Polyakov loop fluctuations and deconfinement in
  the limit of heavy quarks}},\ }\href
  {https://doi.org/10.1103/PhysRevD.90.074035} {\bibfield  {journal} {\bibinfo
  {journal} {Phys. Rev. D}\ }\textbf {\bibinfo {volume} {90}},\ \bibinfo
  {pages} {074035} (\bibinfo {year} {2014})},\ \Eprint
  {https://arxiv.org/abs/1406.4050} {arXiv:1406.4050 [hep-ph]} \BibitemShut
  {NoStop}%
\bibitem [{\citenamefont {Fukushima}(2004)}]{Fukushima:2003fw}%
  \BibitemOpen
  \bibfield  {author} {\bibinfo {author} {\bibfnamefont {K.}~\bibnamefont
  {Fukushima}},\ }\bibfield  {title} {\bibinfo {title} {{Chiral effective model
  with the Polyakov loop}},\ }\href
  {https://doi.org/10.1016/j.physletb.2004.04.027} {\bibfield  {journal}
  {\bibinfo  {journal} {Phys. Lett. B}\ }\textbf {\bibinfo {volume} {591}},\
  \bibinfo {pages} {277} (\bibinfo {year} {2004})},\ \Eprint
  {https://arxiv.org/abs/hep-ph/0310121} {arXiv:hep-ph/0310121} \BibitemShut
  {NoStop}%
\bibitem [{\citenamefont {Zwanziger}(2005)}]{Zwanziger:2004np}%
  \BibitemOpen
  \bibfield  {author} {\bibinfo {author} {\bibfnamefont {D.}~\bibnamefont
  {Zwanziger}},\ }\bibfield  {title} {\bibinfo {title} {{Equation of state of
  gluon plasma from fundamental modular region}},\ }\href
  {https://doi.org/10.1103/PhysRevLett.94.182301} {\bibfield  {journal}
  {\bibinfo  {journal} {Phys. Rev. Lett.}\ }\textbf {\bibinfo {volume} {94}},\
  \bibinfo {pages} {182301} (\bibinfo {year} {2005})},\ \Eprint
  {https://arxiv.org/abs/hep-ph/0407103} {arXiv:hep-ph/0407103} \BibitemShut
  {NoStop}%
\bibitem [{\citenamefont {Szczepaniak}\ \emph {et~al.}(1996)\citenamefont
  {Szczepaniak}, \citenamefont {Swanson}, \citenamefont {Ji},\ and\
  \citenamefont {Cotanch}}]{eric_gb}%
  \BibitemOpen
  \bibfield  {author} {\bibinfo {author} {\bibfnamefont {A.}~\bibnamefont
  {Szczepaniak}}, \bibinfo {author} {\bibfnamefont {E.~S.}\ \bibnamefont
  {Swanson}}, \bibinfo {author} {\bibfnamefont {C.-R.}\ \bibnamefont {Ji}},\
  and\ \bibinfo {author} {\bibfnamefont {S.~R.}\ \bibnamefont {Cotanch}},\
  }\bibfield  {title} {\bibinfo {title} {{Glueball spectroscopy in a
  relativistic many body approach to hadron structure}},\ }\href
  {https://doi.org/10.1103/PhysRevLett.76.2011} {\bibfield  {journal} {\bibinfo
   {journal} {Phys. Rev. Lett.}\ }\textbf {\bibinfo {volume} {76}},\ \bibinfo
  {pages} {2011} (\bibinfo {year} {1996})},\ \Eprint
  {https://arxiv.org/abs/hep-ph/9511422} {arXiv:hep-ph/9511422} \BibitemShut
  {NoStop}%
\bibitem [{\citenamefont {Georgi}(1982)}]{Georgi:1982jb}%
  \BibitemOpen
  \bibfield  {author} {\bibinfo {author} {\bibfnamefont {H.}~\bibnamefont
  {Georgi}},\ }\href {https://doi.org/10.1201/9780429499210} {\emph {\bibinfo
  {title} {{Lie Algebras in Particle Physics. From Isospin to Unified
  Theories}}}},\ Vol.~\bibinfo {volume} {54}\ (\bibinfo  {publisher} {CRC
  Press, UK},\ \bibinfo {year} {1982})\BibitemShut {NoStop}%
\bibitem [{\citenamefont {Gross}\ \emph {et~al.}(1981)\citenamefont {Gross},
  \citenamefont {Pisarski},\ and\ \citenamefont {Yaffe}}]{instanton}%
  \BibitemOpen
  \bibfield  {author} {\bibinfo {author} {\bibfnamefont {D.~J.}\ \bibnamefont
  {Gross}}, \bibinfo {author} {\bibfnamefont {R.~D.}\ \bibnamefont
  {Pisarski}},\ and\ \bibinfo {author} {\bibfnamefont {L.~G.}\ \bibnamefont
  {Yaffe}},\ }\bibfield  {title} {\bibinfo {title} {{QCD and Instantons at
  Finite Temperature}},\ }\href {https://doi.org/10.1103/RevModPhys.53.43}
  {\bibfield  {journal} {\bibinfo  {journal} {Rev. Mod. Phys.}\ }\textbf
  {\bibinfo {volume} {53}},\ \bibinfo {pages} {43} (\bibinfo {year}
  {1981})}\BibitemShut {NoStop}%
\bibitem [{\citenamefont {Dumitru}\ \emph {et~al.}(2014)\citenamefont
  {Dumitru}, \citenamefont {Guo},\ and\ \citenamefont
  {Korthals~Altes}}]{Dumitru:2013xna}%
  \BibitemOpen
  \bibfield  {author} {\bibinfo {author} {\bibfnamefont {A.}~\bibnamefont
  {Dumitru}}, \bibinfo {author} {\bibfnamefont {Y.}~\bibnamefont {Guo}},\ and\
  \bibinfo {author} {\bibfnamefont {C.~P.}\ \bibnamefont {Korthals~Altes}},\
  }\bibfield  {title} {\bibinfo {title} {{Two-loop perturbative corrections to
  the thermal effective potential in gluodynamics}},\ }\href
  {https://doi.org/10.1103/PhysRevD.89.016009} {\bibfield  {journal} {\bibinfo
  {journal} {Phys. Rev. D}\ }\textbf {\bibinfo {volume} {89}},\ \bibinfo
  {pages} {016009} (\bibinfo {year} {2014})},\ \Eprint
  {https://arxiv.org/abs/1305.6846} {arXiv:1305.6846 [hep-ph]} \BibitemShut
  {NoStop}%
\bibitem [{\citenamefont {Weiss}(1981)}]{Weiss:1980rj}%
  \BibitemOpen
  \bibfield  {author} {\bibinfo {author} {\bibfnamefont {N.}~\bibnamefont
  {Weiss}},\ }\bibfield  {title} {\bibinfo {title} {{The Effective Potential
  for the Order Parameter of Gauge Theories at Finite Temperature}},\ }\href
  {https://doi.org/10.1103/PhysRevD.24.475} {\bibfield  {journal} {\bibinfo
  {journal} {Phys. Rev. D}\ }\textbf {\bibinfo {volume} {24}},\ \bibinfo
  {pages} {475} (\bibinfo {year} {1981})}\BibitemShut {NoStop}%
\bibitem [{\citenamefont {Gocksch}\ and\ \citenamefont
  {Pisarski}(1993)}]{Gocksch:1993iy}%
  \BibitemOpen
  \bibfield  {author} {\bibinfo {author} {\bibfnamefont {A.}~\bibnamefont
  {Gocksch}}\ and\ \bibinfo {author} {\bibfnamefont {R.~D.}\ \bibnamefont
  {Pisarski}},\ }\bibfield  {title} {\bibinfo {title} {{Partition function for
  the eigenvalues of the Wilson line}},\ }\href
  {https://doi.org/10.1016/0550-3213(93)90123-7} {\bibfield  {journal}
  {\bibinfo  {journal} {Nucl. Phys. B}\ }\textbf {\bibinfo {volume} {402}},\
  \bibinfo {pages} {657} (\bibinfo {year} {1993})},\ \Eprint
  {https://arxiv.org/abs/hep-ph/9302233} {arXiv:hep-ph/9302233} \BibitemShut
  {NoStop}%
\bibitem [{\citenamefont {von Smekal}\ \emph {et~al.}(1998)\citenamefont {von
  Smekal}, \citenamefont {Hauck},\ and\ \citenamefont
  {Alkofer}}]{vonSmekal:1997ern}%
  \BibitemOpen
  \bibfield  {author} {\bibinfo {author} {\bibfnamefont {L.}~\bibnamefont {von
  Smekal}}, \bibinfo {author} {\bibfnamefont {A.}~\bibnamefont {Hauck}},\ and\
  \bibinfo {author} {\bibfnamefont {R.}~\bibnamefont {Alkofer}},\ }\bibfield
  {title} {\bibinfo {title} {{A Solution to Coupled Dyson\textendash{}Schwinger
  Equations for Gluons and Ghosts in Landau Gauge}},\ }\href
  {https://doi.org/10.1006/aphy.1998.5806} {\bibfield  {journal} {\bibinfo
  {journal} {Annals Phys.}\ }\textbf {\bibinfo {volume} {267}},\ \bibinfo
  {pages} {1} (\bibinfo {year} {1998})},\ \bibinfo {note} {[Erratum: Annals
  Phys. 269, 182 (1998)]},\ \Eprint {https://arxiv.org/abs/hep-ph/9707327}
  {arXiv:hep-ph/9707327} \BibitemShut {NoStop}%
\bibitem [{\citenamefont {von Smekal}\ \emph {et~al.}(1997)\citenamefont {von
  Smekal}, \citenamefont {Alkofer},\ and\ \citenamefont
  {Hauck}}]{vonSmekal:1997ohs}%
  \BibitemOpen
  \bibfield  {author} {\bibinfo {author} {\bibfnamefont {L.}~\bibnamefont {von
  Smekal}}, \bibinfo {author} {\bibfnamefont {R.}~\bibnamefont {Alkofer}},\
  and\ \bibinfo {author} {\bibfnamefont {A.}~\bibnamefont {Hauck}},\ }\bibfield
   {title} {\bibinfo {title} {{The Infrared behavior of gluon and ghost
  propagators in Landau gauge QCD}},\ }\href
  {https://doi.org/10.1103/PhysRevLett.79.3591} {\bibfield  {journal} {\bibinfo
   {journal} {Phys. Rev. Lett.}\ }\textbf {\bibinfo {volume} {79}},\ \bibinfo
  {pages} {3591} (\bibinfo {year} {1997})},\ \Eprint
  {https://arxiv.org/abs/hep-ph/9705242} {arXiv:hep-ph/9705242} \BibitemShut
  {NoStop}%
\bibitem [{\citenamefont {Fischer}\ \emph {et~al.}(2009)\citenamefont
  {Fischer}, \citenamefont {Maas},\ and\ \citenamefont
  {Pawlowski}}]{Fischer:2008uz}%
  \BibitemOpen
  \bibfield  {author} {\bibinfo {author} {\bibfnamefont {C.~S.}\ \bibnamefont
  {Fischer}}, \bibinfo {author} {\bibfnamefont {A.}~\bibnamefont {Maas}},\ and\
  \bibinfo {author} {\bibfnamefont {J.~M.}\ \bibnamefont {Pawlowski}},\
  }\bibfield  {title} {\bibinfo {title} {{On the infrared behavior of Landau
  gauge Yang-Mills theory}},\ }\href
  {https://doi.org/10.1016/j.aop.2009.07.009} {\bibfield  {journal} {\bibinfo
  {journal} {Annals Phys.}\ }\textbf {\bibinfo {volume} {324}},\ \bibinfo
  {pages} {2408} (\bibinfo {year} {2009})},\ \Eprint
  {https://arxiv.org/abs/0810.1987} {arXiv:0810.1987 [hep-ph]} \BibitemShut
  {NoStop}%
\bibitem [{\citenamefont {Aguilar}\ \emph {et~al.}(2008)\citenamefont
  {Aguilar}, \citenamefont {Binosi},\ and\ \citenamefont
  {Papavassiliou}}]{Aguilar:2008xm}%
  \BibitemOpen
  \bibfield  {author} {\bibinfo {author} {\bibfnamefont {A.~C.}\ \bibnamefont
  {Aguilar}}, \bibinfo {author} {\bibfnamefont {D.}~\bibnamefont {Binosi}},\
  and\ \bibinfo {author} {\bibfnamefont {J.}~\bibnamefont {Papavassiliou}},\
  }\bibfield  {title} {\bibinfo {title} {{Gluon and ghost propagators in the
  Landau gauge: Deriving lattice results from Schwinger-Dyson equations}},\
  }\href {https://doi.org/10.1103/PhysRevD.78.025010} {\bibfield  {journal}
  {\bibinfo  {journal} {Phys. Rev. D}\ }\textbf {\bibinfo {volume} {78}},\
  \bibinfo {pages} {025010} (\bibinfo {year} {2008})},\ \Eprint
  {https://arxiv.org/abs/0802.1870} {arXiv:0802.1870 [hep-ph]} \BibitemShut
  {NoStop}%
\bibitem [{\citenamefont {Fister}\ and\ \citenamefont
  {Pawlowski}(2013)}]{referee1}%
  \BibitemOpen
  \bibfield  {author} {\bibinfo {author} {\bibfnamefont {L.}~\bibnamefont
  {Fister}}\ and\ \bibinfo {author} {\bibfnamefont {J.~M.}\ \bibnamefont
  {Pawlowski}},\ }\bibfield  {title} {\bibinfo {title} {{Confinement from
  Correlation Functions}},\ }\href {https://doi.org/10.1103/PhysRevD.88.045010}
  {\bibfield  {journal} {\bibinfo  {journal} {Phys. Rev. D}\ }\textbf {\bibinfo
  {volume} {88}},\ \bibinfo {pages} {045010} (\bibinfo {year} {2013})},\
  \Eprint {https://arxiv.org/abs/1301.4163} {arXiv:1301.4163 [hep-ph]}
  \BibitemShut {NoStop}%
\bibitem [{\citenamefont {Fischer}\ \emph {et~al.}(2014)\citenamefont
  {Fischer}, \citenamefont {Luecker},\ and\ \citenamefont
  {Welzbacher}}]{referee2}%
  \BibitemOpen
  \bibfield  {author} {\bibinfo {author} {\bibfnamefont {C.~S.}\ \bibnamefont
  {Fischer}}, \bibinfo {author} {\bibfnamefont {J.}~\bibnamefont {Luecker}},\
  and\ \bibinfo {author} {\bibfnamefont {C.~A.}\ \bibnamefont {Welzbacher}},\
  }\bibfield  {title} {\bibinfo {title} {{Phase structure of three and four
  flavor QCD}},\ }\href {https://doi.org/10.1103/PhysRevD.90.034022} {\bibfield
   {journal} {\bibinfo  {journal} {Phys. Rev. D}\ }\textbf {\bibinfo {volume}
  {90}},\ \bibinfo {pages} {034022} (\bibinfo {year} {2014})},\ \Eprint
  {https://arxiv.org/abs/1405.4762} {arXiv:1405.4762 [hep-ph]} \BibitemShut
  {NoStop}%
\bibitem [{\citenamefont {Cyrol}\ \emph {et~al.}(2018)\citenamefont {Cyrol},
  \citenamefont {Mitter}, \citenamefont {Pawlowski},\ and\ \citenamefont
  {Strodthoff}}]{referee3}%
  \BibitemOpen
  \bibfield  {author} {\bibinfo {author} {\bibfnamefont {A.~K.}\ \bibnamefont
  {Cyrol}}, \bibinfo {author} {\bibfnamefont {M.}~\bibnamefont {Mitter}},
  \bibinfo {author} {\bibfnamefont {J.~M.}\ \bibnamefont {Pawlowski}},\ and\
  \bibinfo {author} {\bibfnamefont {N.}~\bibnamefont {Strodthoff}},\ }\bibfield
   {title} {\bibinfo {title} {{Nonperturbative finite-temperature Yang-Mills
  theory}},\ }\href {https://doi.org/10.1103/PhysRevD.97.054015} {\bibfield
  {journal} {\bibinfo  {journal} {Phys. Rev. D}\ }\textbf {\bibinfo {volume}
  {97}},\ \bibinfo {pages} {054015} (\bibinfo {year} {2018})},\ \Eprint
  {https://arxiv.org/abs/1708.03482} {arXiv:1708.03482 [hep-ph]} \BibitemShut
  {NoStop}%
\bibitem [{\citenamefont {Maas}(2013)}]{Maas:2011se}%
  \BibitemOpen
  \bibfield  {author} {\bibinfo {author} {\bibfnamefont {A.}~\bibnamefont
  {Maas}},\ }\bibfield  {title} {\bibinfo {title} {{Describing gauge bosons at
  zero and finite temperature}},\ }\href
  {https://doi.org/10.1016/j.physrep.2012.11.002} {\bibfield  {journal}
  {\bibinfo  {journal} {Phys. Rept.}\ }\textbf {\bibinfo {volume} {524}},\
  \bibinfo {pages} {203} (\bibinfo {year} {2013})},\ \Eprint
  {https://arxiv.org/abs/1106.3942} {arXiv:1106.3942 [hep-ph]} \BibitemShut
  {NoStop}%
\bibitem [{\citenamefont {Lo}\ and\ \citenamefont {Swanson}(2010)}]{Lo:2009ud}%
  \BibitemOpen
  \bibfield  {author} {\bibinfo {author} {\bibfnamefont {P.~M.}\ \bibnamefont
  {Lo}}\ and\ \bibinfo {author} {\bibfnamefont {E.~S.}\ \bibnamefont
  {Swanson}},\ }\bibfield  {title} {\bibinfo {title} {{Confinement Models at
  Finite Temperature and Density}},\ }\href
  {https://doi.org/10.1103/PhysRevD.81.034030} {\bibfield  {journal} {\bibinfo
  {journal} {Phys. Rev. D}\ }\textbf {\bibinfo {volume} {81}},\ \bibinfo
  {pages} {034030} (\bibinfo {year} {2010})},\ \Eprint
  {https://arxiv.org/abs/0908.4099} {arXiv:0908.4099 [hep-ph]} \BibitemShut
  {NoStop}%
\bibitem [{\citenamefont {Bernard}(1974)}]{Bernard}%
  \BibitemOpen
  \bibfield  {author} {\bibinfo {author} {\bibfnamefont {C.~W.}\ \bibnamefont
  {Bernard}},\ }\bibfield  {title} {\bibinfo {title} {Feynman rules for gauge
  theories at finite temperature},\ }\href
  {https://doi.org/10.1103/PhysRevD.9.3312} {\bibfield  {journal} {\bibinfo
  {journal} {Phys. Rev. D}\ }\textbf {\bibinfo {volume} {9}},\ \bibinfo {pages}
  {3312} (\bibinfo {year} {1974})}\BibitemShut {NoStop}%
\bibitem [{\citenamefont {Bogolubsky}\ \emph {et~al.}(2009)\citenamefont
  {Bogolubsky}, \citenamefont {Ilgenfritz}, \citenamefont {Muller-Preussker},\
  and\ \citenamefont {Sternbeck}}]{Bogolubsky:2009dc}%
  \BibitemOpen
  \bibfield  {author} {\bibinfo {author} {\bibfnamefont {I.}~\bibnamefont
  {Bogolubsky}}, \bibinfo {author} {\bibfnamefont {E.}~\bibnamefont
  {Ilgenfritz}}, \bibinfo {author} {\bibfnamefont {M.}~\bibnamefont
  {Muller-Preussker}},\ and\ \bibinfo {author} {\bibfnamefont {A.}~\bibnamefont
  {Sternbeck}},\ }\bibfield  {title} {\bibinfo {title} {{Lattice gluodynamics
  computation of Landau gauge Green's functions in the deep infrared}},\ }\href
  {https://doi.org/10.1016/j.physletb.2009.04.076} {\bibfield  {journal}
  {\bibinfo  {journal} {Phys. Lett. B}\ }\textbf {\bibinfo {volume} {676}},\
  \bibinfo {pages} {69} (\bibinfo {year} {2009})},\ \Eprint
  {https://arxiv.org/abs/0901.0736} {arXiv:0901.0736 [hep-lat]} \BibitemShut
  {NoStop}%
\bibitem [{\citenamefont {Alkofer}\ and\ \citenamefont {von
  Smekal}(2001)}]{Alkofer:2000wg}%
  \BibitemOpen
  \bibfield  {author} {\bibinfo {author} {\bibfnamefont {R.}~\bibnamefont
  {Alkofer}}\ and\ \bibinfo {author} {\bibfnamefont {L.}~\bibnamefont {von
  Smekal}},\ }\bibfield  {title} {\bibinfo {title} {{The Infrared behavior of
  QCD Green's functions: Confinement dynamical symmetry breaking, and hadrons
  as relativistic bound states}},\ }\href
  {https://doi.org/10.1016/S0370-1573(01)00010-2} {\bibfield  {journal}
  {\bibinfo  {journal} {Phys. Rept.}\ }\textbf {\bibinfo {volume} {353}},\
  \bibinfo {pages} {281} (\bibinfo {year} {2001})},\ \Eprint
  {https://arxiv.org/abs/hep-ph/0007355} {arXiv:hep-ph/0007355} \BibitemShut
  {NoStop}%
\bibitem [{\citenamefont {Iritani}\ \emph {et~al.}(2009)\citenamefont
  {Iritani}, \citenamefont {Suganuma},\ and\ \citenamefont
  {Iida}}]{Iritani:2009mp}%
  \BibitemOpen
  \bibfield  {author} {\bibinfo {author} {\bibfnamefont {T.}~\bibnamefont
  {Iritani}}, \bibinfo {author} {\bibfnamefont {H.}~\bibnamefont {Suganuma}},\
  and\ \bibinfo {author} {\bibfnamefont {H.}~\bibnamefont {Iida}},\ }\bibfield
  {title} {\bibinfo {title} {{Gluon-propagator functional form in the Landau
  gauge in SU(3) lattice QCD: Yukawa-type gluon propagator and anomalous gluon
  spectral function}},\ }\href {https://doi.org/10.1103/PhysRevD.80.114505}
  {\bibfield  {journal} {\bibinfo  {journal} {Phys. Rev. D}\ }\textbf {\bibinfo
  {volume} {80}},\ \bibinfo {pages} {114505} (\bibinfo {year} {2009})},\
  \Eprint {https://arxiv.org/abs/0908.1311} {arXiv:0908.1311 [hep-lat]}
  \BibitemShut {NoStop}%
\bibitem [{\citenamefont {Maas}(2017)}]{Maas:2017csm}%
  \BibitemOpen
  \bibfield  {author} {\bibinfo {author} {\bibfnamefont {A.}~\bibnamefont
  {Maas}},\ }\bibfield  {title} {\bibinfo {title} {{Dependence of the
  propagators on the sampling of Gribov copies inside the first Gribov region
  of Landau gauge}},\ }\href {https://doi.org/10.1016/j.aop.2017.10.003}
  {\bibfield  {journal} {\bibinfo  {journal} {Annals Phys.}\ }\textbf {\bibinfo
  {volume} {387}},\ \bibinfo {pages} {29} (\bibinfo {year} {2017})},\ \Eprint
  {https://arxiv.org/abs/1705.03812} {arXiv:1705.03812 [hep-lat]} \BibitemShut
  {NoStop}%
\bibitem [{\citenamefont {Maas}(2020)}]{Maas:2019ggf}%
  \BibitemOpen
  \bibfield  {author} {\bibinfo {author} {\bibfnamefont {A.}~\bibnamefont
  {Maas}},\ }\bibfield  {title} {\bibinfo {title} {{Constraining the
  gauge-fixed Lagrangian in minimal Landau gauge}},\ }\href
  {https://doi.org/10.21468/SciPostPhys.8.5.071} {\bibfield  {journal}
  {\bibinfo  {journal} {SciPost Phys.}\ }\textbf {\bibinfo {volume} {8}},\
  \bibinfo {pages} {071} (\bibinfo {year} {2020})},\ \Eprint
  {https://arxiv.org/abs/1907.10435} {arXiv:1907.10435 [hep-lat]} \BibitemShut
  {NoStop}%
\bibitem [{\citenamefont {Dudal}\ \emph {et~al.}(2018)\citenamefont {Dudal},
  \citenamefont {Oliveira},\ and\ \citenamefont {Silva}}]{orlando1}%
  \BibitemOpen
  \bibfield  {author} {\bibinfo {author} {\bibfnamefont {D.}~\bibnamefont
  {Dudal}}, \bibinfo {author} {\bibfnamefont {O.}~\bibnamefont {Oliveira}},\
  and\ \bibinfo {author} {\bibfnamefont {P.~J.}\ \bibnamefont {Silva}},\
  }\bibfield  {title} {\bibinfo {title} {{High precision statistical Landau
  gauge lattice gluon propagator computation vs. the
  Gribov\textendash{}Zwanziger approach}},\ }\href
  {https://doi.org/10.1016/j.aop.2018.08.019} {\bibfield  {journal} {\bibinfo
  {journal} {Annals Phys.}\ }\textbf {\bibinfo {volume} {397}},\ \bibinfo
  {pages} {351} (\bibinfo {year} {2018})},\ \Eprint
  {https://arxiv.org/abs/1803.02281} {arXiv:1803.02281 [hep-lat]} \BibitemShut
  {NoStop}%
\bibitem [{\citenamefont {Falc\~ao}\ \emph {et~al.}(2020)\citenamefont
  {Falc\~ao}, \citenamefont {Oliveira},\ and\ \citenamefont
  {Silva}}]{orlando2}%
  \BibitemOpen
  \bibfield  {author} {\bibinfo {author} {\bibfnamefont {A.~F.}\ \bibnamefont
  {Falc\~ao}}, \bibinfo {author} {\bibfnamefont {O.}~\bibnamefont {Oliveira}},\
  and\ \bibinfo {author} {\bibfnamefont {P.~J.}\ \bibnamefont {Silva}},\
  }\bibfield  {title} {\bibinfo {title} {Analytic structure of the lattice
  landau gauge gluon and ghost propagators},\ }\href
  {https://doi.org/10.1103/PhysRevD.102.114518} {\bibfield  {journal} {\bibinfo
   {journal} {Phys. Rev. D}\ }\textbf {\bibinfo {volume} {102}},\ \bibinfo
  {pages} {114518} (\bibinfo {year} {2020})}\BibitemShut {NoStop}%
\bibitem [{\citenamefont {Weiss}(1982)}]{Weiss:1981ev}%
  \BibitemOpen
  \bibfield  {author} {\bibinfo {author} {\bibfnamefont {N.}~\bibnamefont
  {Weiss}},\ }\bibfield  {title} {\bibinfo {title} {{The Wilson Line in Finite
  Temperature Gauge Theories}},\ }\href
  {https://doi.org/10.1103/PhysRevD.25.2667} {\bibfield  {journal} {\bibinfo
  {journal} {Phys. Rev. D}\ }\textbf {\bibinfo {volume} {25}},\ \bibinfo
  {pages} {2667} (\bibinfo {year} {1982})}\BibitemShut {NoStop}%
\bibitem [{\citenamefont {Dumitru}\ and\ \citenamefont
  {Pisarski}(2002)}]{rob_dof}%
  \BibitemOpen
  \bibfield  {author} {\bibinfo {author} {\bibfnamefont {A.}~\bibnamefont
  {Dumitru}}\ and\ \bibinfo {author} {\bibfnamefont {R.~D.}\ \bibnamefont
  {Pisarski}},\ }\bibfield  {title} {\bibinfo {title} {{Degrees of freedom and
  the deconfining phase transition}},\ }\href
  {https://doi.org/10.1016/S0370-2693(01)01424-1} {\bibfield  {journal}
  {\bibinfo  {journal} {Phys. Lett. B}\ }\textbf {\bibinfo {volume} {525}},\
  \bibinfo {pages} {95} (\bibinfo {year} {2002})},\ \Eprint
  {https://arxiv.org/abs/hep-ph/0106176} {arXiv:hep-ph/0106176} \BibitemShut
  {NoStop}%
\bibitem [{\citenamefont {Silva}\ \emph {et~al.}(2014)\citenamefont {Silva},
  \citenamefont {Oliveira}, \citenamefont {Bicudo},\ and\ \citenamefont
  {Cardoso}}]{lqcd_1}%
  \BibitemOpen
  \bibfield  {author} {\bibinfo {author} {\bibfnamefont {P.~J.}\ \bibnamefont
  {Silva}}, \bibinfo {author} {\bibfnamefont {O.}~\bibnamefont {Oliveira}},
  \bibinfo {author} {\bibfnamefont {P.}~\bibnamefont {Bicudo}},\ and\ \bibinfo
  {author} {\bibfnamefont {N.}~\bibnamefont {Cardoso}},\ }\bibfield  {title}
  {\bibinfo {title} {Gluon screening mass at finite temperature from the landau
  gauge gluon propagator in lattice qcd},\ }\href
  {https://doi.org/10.1103/PhysRevD.89.074503} {\bibfield  {journal} {\bibinfo
  {journal} {Phys. Rev. D}\ }\textbf {\bibinfo {volume} {89}},\ \bibinfo
  {pages} {074503} (\bibinfo {year} {2014})}\BibitemShut {NoStop}%
\bibitem [{\citenamefont {Maas}\ \emph {et~al.}(2012)\citenamefont {Maas},
  \citenamefont {Pawlowski}, \citenamefont {von Smekal},\ and\ \citenamefont
  {Spielmann}}]{Maas:2011ez}%
  \BibitemOpen
  \bibfield  {author} {\bibinfo {author} {\bibfnamefont {A.}~\bibnamefont
  {Maas}}, \bibinfo {author} {\bibfnamefont {J.~M.}\ \bibnamefont {Pawlowski}},
  \bibinfo {author} {\bibfnamefont {L.}~\bibnamefont {von Smekal}},\ and\
  \bibinfo {author} {\bibfnamefont {D.}~\bibnamefont {Spielmann}},\ }\bibfield
  {title} {\bibinfo {title} {{The Gluon propagator close to criticality}},\
  }\href {https://doi.org/10.1103/PhysRevD.85.034037} {\bibfield  {journal}
  {\bibinfo  {journal} {Phys. Rev. D}\ }\textbf {\bibinfo {volume} {85}},\
  \bibinfo {pages} {034037} (\bibinfo {year} {2012})},\ \Eprint
  {https://arxiv.org/abs/1110.6340} {arXiv:1110.6340 [hep-lat]} \BibitemShut
  {NoStop}%
\bibitem [{\citenamefont {Reinosa}\ \emph {et~al.}(2016)\citenamefont
  {Reinosa}, \citenamefont {Serreau}, \citenamefont {Tissier},\ and\
  \citenamefont {Wschebor}}]{2loop}%
  \BibitemOpen
  \bibfield  {author} {\bibinfo {author} {\bibfnamefont {U.}~\bibnamefont
  {Reinosa}}, \bibinfo {author} {\bibfnamefont {J.}~\bibnamefont {Serreau}},
  \bibinfo {author} {\bibfnamefont {M.}~\bibnamefont {Tissier}},\ and\ \bibinfo
  {author} {\bibfnamefont {N.}~\bibnamefont {Wschebor}},\ }\bibfield  {title}
  {\bibinfo {title} {{Two-loop study of the deconfinement transition in
  Yang-Mills theories: SU(3) and beyond}},\ }\href
  {https://doi.org/10.1103/PhysRevD.93.105002} {\bibfield  {journal} {\bibinfo
  {journal} {Phys. Rev. D}\ }\textbf {\bibinfo {volume} {93}},\ \bibinfo
  {pages} {105002} (\bibinfo {year} {2016})},\ \Eprint
  {https://arxiv.org/abs/1511.07690} {arXiv:1511.07690 [hep-th]} \BibitemShut
  {NoStop}%
\bibitem [{\citenamefont {Langfeld}\ and\ \citenamefont
  {Moyaerts}(2004)}]{Langfeld:2004qs}%
  \BibitemOpen
  \bibfield  {author} {\bibinfo {author} {\bibfnamefont {K.}~\bibnamefont
  {Langfeld}}\ and\ \bibinfo {author} {\bibfnamefont {L.}~\bibnamefont
  {Moyaerts}},\ }\bibfield  {title} {\bibinfo {title} {{Propagators in Coulomb
  gauge from SU(2) lattice gauge theory}},\ }\href
  {https://doi.org/10.1103/PhysRevD.70.074507} {\bibfield  {journal} {\bibinfo
  {journal} {Phys. Rev. D}\ }\textbf {\bibinfo {volume} {70}},\ \bibinfo
  {pages} {074507} (\bibinfo {year} {2004})},\ \Eprint
  {https://arxiv.org/abs/hep-lat/0406024} {arXiv:hep-lat/0406024} \BibitemShut
  {NoStop}%
\bibitem [{\citenamefont {Dudal}\ \emph {et~al.}(2008)\citenamefont {Dudal},
  \citenamefont {Sorella}, \citenamefont {Vandersickel},\ and\ \citenamefont
  {Verschelde}}]{Dudal:2007cw}%
  \BibitemOpen
  \bibfield  {author} {\bibinfo {author} {\bibfnamefont {D.}~\bibnamefont
  {Dudal}}, \bibinfo {author} {\bibfnamefont {S.~P.}\ \bibnamefont {Sorella}},
  \bibinfo {author} {\bibfnamefont {N.}~\bibnamefont {Vandersickel}},\ and\
  \bibinfo {author} {\bibfnamefont {H.}~\bibnamefont {Verschelde}},\ }\bibfield
   {title} {\bibinfo {title} {{New features of the gluon and ghost propagator
  in the infrared region from the Gribov-Zwanziger approach}},\ }\href
  {https://doi.org/10.1103/PhysRevD.77.071501} {\bibfield  {journal} {\bibinfo
  {journal} {Phys. Rev. D}\ }\textbf {\bibinfo {volume} {77}},\ \bibinfo
  {pages} {071501} (\bibinfo {year} {2008})},\ \Eprint
  {https://arxiv.org/abs/0711.4496} {arXiv:0711.4496 [hep-th]} \BibitemShut
  {NoStop}%
\bibitem [{\citenamefont {Lucini}\ \emph {et~al.}(2012)\citenamefont {Lucini},
  \citenamefont {Rago},\ and\ \citenamefont {Rinaldi}}]{lucini_td}%
  \BibitemOpen
  \bibfield  {author} {\bibinfo {author} {\bibfnamefont {B.}~\bibnamefont
  {Lucini}}, \bibinfo {author} {\bibfnamefont {A.}~\bibnamefont {Rago}},\ and\
  \bibinfo {author} {\bibfnamefont {E.}~\bibnamefont {Rinaldi}},\ }\bibfield
  {title} {\bibinfo {title} {{SU($N_c$) gauge theories at deconfinement}},\
  }\href {https://doi.org/10.1016/j.physletb.2012.04.070} {\bibfield  {journal}
  {\bibinfo  {journal} {Phys. Lett. B}\ }\textbf {\bibinfo {volume} {712}},\
  \bibinfo {pages} {279} (\bibinfo {year} {2012})},\ \Eprint
  {https://arxiv.org/abs/1202.6684} {arXiv:1202.6684 [hep-lat]} \BibitemShut
  {NoStop}%
\bibitem [{\citenamefont {Lucini}\ \emph {et~al.}(2004)\citenamefont {Lucini},
  \citenamefont {Teper},\ and\ \citenamefont {Wenger}}]{lucini_gb}%
  \BibitemOpen
  \bibfield  {author} {\bibinfo {author} {\bibfnamefont {B.}~\bibnamefont
  {Lucini}}, \bibinfo {author} {\bibfnamefont {M.}~\bibnamefont {Teper}},\ and\
  \bibinfo {author} {\bibfnamefont {U.}~\bibnamefont {Wenger}},\ }\bibfield
  {title} {\bibinfo {title} {{Glueballs and k-strings in SU(N) gauge theories:
  Calculations with improved operators}},\ }\href
  {https://doi.org/10.1088/1126-6708/2004/06/012} {\bibfield  {journal}
  {\bibinfo  {journal} {JHEP}\ }\textbf {\bibinfo {volume} {06}},\ \bibinfo
  {pages} {012}},\ \Eprint {https://arxiv.org/abs/hep-lat/0404008}
  {arXiv:hep-lat/0404008} \BibitemShut {NoStop}%
\bibitem [{\citenamefont {Szczepaniak}\ and\ \citenamefont
  {Swanson}(2001)}]{eric_cgauge}%
  \BibitemOpen
  \bibfield  {author} {\bibinfo {author} {\bibfnamefont {A.~P.}\ \bibnamefont
  {Szczepaniak}}\ and\ \bibinfo {author} {\bibfnamefont {E.~S.}\ \bibnamefont
  {Swanson}},\ }\bibfield  {title} {\bibinfo {title} {{Coulomb gauge QCD,
  confinement, and the constituent representation}},\ }\href
  {https://doi.org/10.1103/PhysRevD.65.025012} {\bibfield  {journal} {\bibinfo
  {journal} {Phys. Rev. D}\ }\textbf {\bibinfo {volume} {65}},\ \bibinfo
  {pages} {025012} (\bibinfo {year} {2001})},\ \Eprint
  {https://arxiv.org/abs/hep-ph/0107078} {arXiv:hep-ph/0107078} \BibitemShut
  {NoStop}%
\bibitem [{\citenamefont {Lacroix}\ \emph {et~al.}(2013)\citenamefont
  {Lacroix}, \citenamefont {Semay}, \citenamefont {Cabrera},\ and\
  \citenamefont {Buisseret}}]{Lacroix2013}%
  \BibitemOpen
  \bibfield  {author} {\bibinfo {author} {\bibfnamefont {G.}~\bibnamefont
  {Lacroix}}, \bibinfo {author} {\bibfnamefont {C.}~\bibnamefont {Semay}},
  \bibinfo {author} {\bibfnamefont {D.}~\bibnamefont {Cabrera}},\ and\ \bibinfo
  {author} {\bibfnamefont {F.}~\bibnamefont {Buisseret}},\ }\bibfield  {title}
  {\bibinfo {title} {{Glueballs and the Yang-Mills plasma in a T-matrix
  approach}},\ }\href {https://doi.org/10.1103/PhysRevD.87.054025} {\bibfield
  {journal} {\bibinfo  {journal} {Phys. Rev. D - Part. Fields, Gravit.
  Cosmol.}\ }\textbf {\bibinfo {volume} {87}},\ \bibinfo {pages} {054025}
  (\bibinfo {year} {2013})}\BibitemShut {NoStop}%
\bibitem [{\citenamefont {Aouane}\ \emph {et~al.}(2012)\citenamefont {Aouane},
  \citenamefont {Bornyakov}, \citenamefont {Ilgenfritz}, \citenamefont
  {Mitrjushkin}, \citenamefont {Muller-Preussker},\ and\ \citenamefont
  {Sternbeck}}]{Aouane:2011fv}%
  \BibitemOpen
  \bibfield  {author} {\bibinfo {author} {\bibfnamefont {R.}~\bibnamefont
  {Aouane}}, \bibinfo {author} {\bibfnamefont {V.}~\bibnamefont {Bornyakov}},
  \bibinfo {author} {\bibfnamefont {E.}~\bibnamefont {Ilgenfritz}}, \bibinfo
  {author} {\bibfnamefont {V.}~\bibnamefont {Mitrjushkin}}, \bibinfo {author}
  {\bibfnamefont {M.}~\bibnamefont {Muller-Preussker}},\ and\ \bibinfo {author}
  {\bibfnamefont {A.}~\bibnamefont {Sternbeck}},\ }\bibfield  {title} {\bibinfo
  {title} {{Landau gauge gluon and ghost propagators at finite temperature from
  quenched lattice QCD}},\ }\href {https://doi.org/10.1103/PhysRevD.85.034501}
  {\bibfield  {journal} {\bibinfo  {journal} {Phys. Rev. D}\ }\textbf {\bibinfo
  {volume} {85}},\ \bibinfo {pages} {034501} (\bibinfo {year} {2012})},\
  \Eprint {https://arxiv.org/abs/1108.1735} {arXiv:1108.1735 [hep-lat]}
  \BibitemShut {NoStop}%
\bibitem [{\citenamefont {Bornyakov}\ and\ \citenamefont
  {Mitrjushkin}(2011)}]{Bornyakov:2010nc}%
  \BibitemOpen
  \bibfield  {author} {\bibinfo {author} {\bibfnamefont {V.~G.}\ \bibnamefont
  {Bornyakov}}\ and\ \bibinfo {author} {\bibfnamefont {V.~K.}\ \bibnamefont
  {Mitrjushkin}},\ }\bibfield  {title} {\bibinfo {title} {{SU(2) lattice gluon
  propagators at finite temperatures in the deep infrared region and Gribov
  copy effects}},\ }\href {https://doi.org/10.1103/PhysRevD.84.094503}
  {\bibfield  {journal} {\bibinfo  {journal} {Phys. Rev. D}\ }\textbf {\bibinfo
  {volume} {84}},\ \bibinfo {pages} {094503} (\bibinfo {year} {2011})},\
  \Eprint {https://arxiv.org/abs/1011.4790} {arXiv:1011.4790 [hep-lat]}
  \BibitemShut {NoStop}%
\bibitem [{\citenamefont {Bicudo}\ \emph {et~al.}(2019)\citenamefont {Bicudo},
  \citenamefont {Cardoso},\ and\ \citenamefont {Cardoso}}]{Bicudo:2017uyy}%
  \BibitemOpen
  \bibfield  {author} {\bibinfo {author} {\bibfnamefont {P.}~\bibnamefont
  {Bicudo}}, \bibinfo {author} {\bibfnamefont {N.}~\bibnamefont {Cardoso}},\
  and\ \bibinfo {author} {\bibfnamefont {M.}~\bibnamefont {Cardoso}},\
  }\bibfield  {title} {\bibinfo {title} {{Pure gauge QCD flux tubes and their
  widths at finite temperature}},\ }\href
  {https://doi.org/10.1016/j.nuclphysb.2019.01.012} {\bibfield  {journal}
  {\bibinfo  {journal} {Nucl. Phys. B}\ }\textbf {\bibinfo {volume} {940}},\
  \bibinfo {pages} {88} (\bibinfo {year} {2019})},\ \Eprint
  {https://arxiv.org/abs/1702.03454} {arXiv:1702.03454 [hep-lat]} \BibitemShut
  {NoStop}%
\end{thebibliography}%


\end{document}